\documentclass[final]{iopart}
\usepackage{iopams}
\usepackage{graphicx,bm,amsfonts,amssymb,amsbsy,texdraw}
\usepackage[active]{srcltx}
\usepackage{dcolumn}
\newcolumntype{.}{D{.}{.}{-1}}
\usepackage[dvips]{hyperref}
\eqnobysec
\newcommand{\sigb}{\mathring{\sigma}}
\newcommand{\tb}{\mathring{\tau}}
\newcommand{\rhob}{\mathring{\rho}}

\newcommand{\eqref}[1]{(\ref{#1})}

\date{\today}
\begin{document}

\title[Fluctuation-induced forces in strongly anisotropic critical systems]{Fluctuation-induced forces in strongly anisotropic critical systems } 
\author{Matthias Burgsm{\"u}ller\dag, H W Diehl\dag\ and M A Shpot\dag\ddag} 
\address{\dag\ Fakult\"at f\"ur Physik, Universit{\"a}t Duisburg-Essen, 47048 Duisburg, Germany}
\address{\ddag\ Institute for Condensed Matter Physics, 79011 Lviv, Ukraine}

\begin{abstract}
 Strongly anisotropic critical systems are considered in a $d$-dimensional film
 geometry. Such systems involve two (or more) distinct correlation
 lengths $\xi_\beta$ and $\xi_\alpha$ that scale as nontrivial powers
 of each other, i.e.\ $\xi_\alpha\sim\xi_\beta^\theta$ with anisotropy index $\theta\ne 1$.
 Thus two fundamental orientations, perpendicular ($\perp$) and parallel ($\|$),
 for which the surface normal is oriented along an $\alpha$- and $\beta$-direction, respectively, must be distinguished. The confinement of critical
 fluctuations caused by the film's boundary planes is shown to induce
 effective forces $\mathcal{F}_C$ that decay as $\mathcal{F}_C\propto
 -(\partial/\partial L)\Delta_{\perp,\|}\,L^{-\zeta_{\perp,\|}}$ as
 the film thickness $L$ becomes large, where the proportionality constants involve nonuniversal amplitudes. The decay exponents $\zeta_{\perp,\|}$ and the Casimir amplitudes $\Delta_{\perp,\|}$ are universal but depend on the type of orientation.  To corroborate these findings,
 $n$-vector models with an $m$-axial bulk Lifshitz point are
 investigated by means of RG methods below the upper critical
 dimension $d^*(m)=4+m/2$ under periodic boundary conditions (PBC) and free BC of an asymptotic form pertaining to the respective ordinary surface transitions. The
 exponents $\zeta_{\perp,\|}$ are determined, and explicit results
 to one- or two-loop order are presented for several Casimir
 amplitudes $\Delta^{\mathrm{BC}}_{\perp,\|}$. The large-$n$ limits of the Casimir amplitudes $\Delta_{\|}^{\mathrm{BC}}/n$ for periodic and Dirichlet BC are shown to be proportional to their critical-point analogues at dimension $d-m/2$. The limiting values $\Delta^{\mathrm{PBC}}_{\|,\perp,\infty}=\lim_{n\to\infty}\Delta_{\|,\perp}^{\mathrm{PBC}}/n$ are determined exactly for the uniaxial cases $(d,m)=(3,1)$ under periodic BC. Unlike $\Delta^{\mathrm{PBC}}_{\|,\infty}$, the amplitude $\Delta^{\mathrm{PBC}}_{\perp,\infty}$ is positive, so that the corresponding Casimir force is repulsive.
\end{abstract}

\section{Introduction}
\label{sec:intro}
When critical fluctuations in a medium are confined by the presence of macroscopic bodies such as walls, long-range effective forces between these bodies are induced. This phenomenon, first pointed out by Fisher and de Gennes \cite{FdG78}, is the thermal analogue of the familiar Casimir effect between metallic objects caused by fluctuations of the electromagnetic field \cite{Cas48}. It has attracted considerable attention during the past 15 years \cite{Kre94,BDT00}. Its indirect experimental verification through the thinning of $^4$He wetting layers near the lambda transition \cite{GC99,GC02, GSGC06} some time ago and the more recent first direct measurement of such Casimir forces for binary fluid mixtures \cite{HHGDB08, GMHNB09,Gam09} are likely to spur further interest and increasing research activities in this field.

Previous studies of critical Casimir forces in statistical physics have focused exclusively on macroscopic media exhibiting \emph{isotropic} scale invariance in the absence of confining walls and macroscopic bodies. The characteristic feature of such systems is that they become self-similar when distances $\Delta x$ along arbitrary directions are rescaled by a scale factor $\ell=\Delta x/\Delta x'$. Alternatively, one can say that the bulk correlation lengths characterizing the decay of correlations along all directions exhibit the same power-law divergence $\sim |T/T_c -1|^{-\nu}$ as the temperature $T$ approaches the bulk critical temperature $T_c$.

In this paper we shall be concerned with \emph{strongly anisotropic} scale-invariant systems. In bulk systems of this kind there exist one or several  principal directions along which coordinate separations $\Delta x_\alpha$ must be rescaled by a nontrivial power $\ell^\theta$ of the scale factor $\ell=\Delta x_\beta/\Delta x_\beta'$ associated with the remaining principal directions in order to have self-similarity of the initial and transformed systems. The asymptotic power laws
\begin{equation}
  \label{eq:xialphabeta}
  \xi_{\alpha,\beta}\sim |T/T_c
-1|^{-\nu_{\alpha,\beta}}
\end{equation}
of the respective bulk correlation lengths $\xi_{\alpha}$ and $\xi_{\beta}$ are governed by distinct exponents $\nu_\alpha=\theta\nu_\beta $ and $\nu_\beta$. Such systems are ubiquitous in nature. Important examples of equilibrium systems whose (multi)critical equilibrium states exhibit anisotropic scale invariance (ASI) are systems at Lifshitz points  \cite{Hor80,Sel92,Die02} and liquid crystals \cite{CL76,Sin00}. Furthermore, ASI is a common feature of the stationary states of many nonequilibrium systems \cite{SZ95}. We shall show that the Casimir forces induced by confinement of fluctuations in systems exhibiting ASI differ \emph{qualitatively and quantitatively} from their analogues for isotropic scale-invariant systems.

Consider a system confined by two parallel planes at a distance $L$, each of which has area $A$. The reduced free energy  of the system per area $A$ can be written as
\begin{equation}\label{eq:Fdec}
 \frac{F}{k_BTA}\approx Lf_{\mathrm{b}}(T,\ldots)+f_{\mathrm{s}}(T,\ldots)+f_{{\rm res}}(L;T,\dots)
  \end{equation} 
in the limit $A\to\infty$ at fixed $L$, where $f_{\mathrm{b}}$ and $f_{\mathrm{s}}$ are $L$-independent bulk and surface excess densities. The ellipsis in $f_{\mathrm{b}}(T,\ldots)$ stands for additional thermodynamic bulk fields (such as magnetic field or chemical potential); the one in $f_{\mathrm{s}}(T, \ldots)$  represents both bulk fields of this kind as well as  additional surface variables (such as surface interaction constants). The $L$-dependence resides in the so-defined residual free-energy density $f_{\rm res}(L;T,\ldots)$. For given medium and boundary planes, this function decays at the bulk critical point as
\begin{equation}
  \label{eq:fres}\label{eq:frescrit}
 f_{{\rm res,crit}}(L)\approx \Delta^{\mathrm{BC}}A_1^{-1}\,(L/L_{1})^{-\zeta}
\end{equation}
as $L\to\infty$, where $A_1$ and $L_1$ are metric factors (units of area and length, respectively). At conventional critical points exhibiting isotropic scale invariance, one has $\zeta=d-1$ for the decay exponent. Furthermore, the metric factors are chosen to satisfy $A_1=L_1^{d-1}$ for systems whose correlation regimes are hyperspherical (i.e., whose bulk correlation lengths $\xi\approx \xi_0 |T/T_c-1|^{-\nu}$ characterizing the decay of correlations along different spatial directions diverge with the same exponent \emph{and} have equal amplitudes $\xi_0$). This choice guarantees that $A_1$ and $L_1$  drop out of equation~\eqref{eq:frescrit}. 

The ``Casimir amplitude'' $\Delta^{\rm BC}$ depends on gross properties of the medium (universality class) and the boundaries, namely, the
boundary conditions (BC) that hold on large length scales \cite{SD08}, but is independent of microscopic details (universal). To avoid confusion, let us explain how the metric factors $A_1$ and $L_1$ ought to be chosen when the hypersphericity condition on the correlation regime is violated in the way weakly anisotropic critical systems do \cite{Doh08}.  For those, the bulk correlation lengths $\xi$ associated with distinct directions diverge with the same critical exponent $\nu$ but involve several different amplitudes $\xi_0$, so that their correlation regime is hyperellipsoidal. Their critical behavior can be expressed in terms of that of standard isotropic systems such as the conventional $\phi^4$ model \cite{Doh08,DC09}. The required transformation that makes the correlation regime hyperspherical --- a similarity transformation followed by a rescaling of the principal radii --- generally changes $A_1$ and $L_1$ into transformed values $A_1'$ and $L_1'$ (see  p.~15--17 of \cite{DC09}]). To define $\Delta^{\mathrm{BC}}$ for weakly anisotropic systems in a consistent manner, one must simply choose $A_1'$ and $L_1'$ according to our rule for isotropic systems,  requiring them to drop out of the corresponding analogue of equation~\eqref{eq:frescrit}.

The anisotropy one encounters in  weakly anisotropic critical systems is of a fairly harmless kind: it can be transformed away, absorbed by a proper choice of  (nonuniversal) coordinates. This is not the case for 
systems exhibiting ASI. Their anisotropy is of a more profound type. This has important consequences for fluctuation-induced forces. General aspects of the orientation of the boundary planes matter. Two fundamentally distinct orientations must be distinguished: parallel ($\|$), for which the normals to the boundary planes are oriented along a $\beta$-direction, and perpendicular ($\perp$), for which the normals are parallel to an $\alpha$- but perpendicular to all $\beta$-directions. If one has $m$ $\alpha$- and $d-m$
$\beta$-directions in a $d$-dimensional system, then the corresponding
decay exponents are given by
\begin{equation}
  \label{eq:zetas}
\zeta_\|=d-m+\theta\, m-1\quad\quad\mbox{and}\quad\quad
\zeta_\perp=(d-m)/\theta+m-1\;.
\end{equation}
In the cases of $m$-axial Lifshitz points, which we explicitly
consider in the following, the value of the anisotropy exponent $\theta$
is close to $1/2$; one has $\theta=1/2+\Or(\epsilon^2)$,
\cite{DS00a,SD01} where $\epsilon\equiv d^*(m)-d$ is
the deviation of the bulk dimension $d$  from the corresponding upper critical
dimension $d^*(m)=4+m/2$ of the system.

The associated Casimir amplitudes $\Delta_\|^{\rm BC}$ and
$\Delta_\perp^{\rm BC}$ depend on the BC, and
 generally differ for perpendicular and parallel
orientations. How they should be defined in order to avoid trivial dependences on nonuniversal metric factors needs to be clarified. The results~\eqref{eq:zetas} for the decay
exponents can be understood by simple scaling
arguments. The residual reduced free energy density $f_{\rm res}(L)$ has
dimension $1/A$ where $A$ scales as $\xi_\alpha^m\xi_\beta^{d-m-1}$
and $\xi_\alpha^{m-1}\xi_\beta^{d-m}$ for parallel and perpendicular
surface orientations, respectively. If $\xi_\alpha$ and
$\xi_\beta$ are the only relevant lengths besides $L$, then
equation~\eqref{eq:zetas} should hold.\footnote{Note that since $\xi_\alpha$ and $\xi_\beta$ are  bulk correlation lengths, they diverge at the bulk critical point. In later sections we shall also consider finite-size correlation lengths. These remain finite at the bulk critical point when $L<\infty$.}

To substantiate these claims and verify explicitly that the Casimir amplitudes differ for parallel and perpendicular slab orientations, we shall investigate a familiar class of prototype $n$-vector models exhibiting ASI --- namely, $O(n)$ $\phi^4$~models with an $m$-axial bulk Lifshitz point (LP) \cite{Hor80,Sel92,Die02} in a slab geometry.
The cases of parallel and perpendicular orientations of the boundary planes will both be studied under periodic (PBC) and free (FBC) boundary conditions. However, when considering FBC, we shall restrict ourselves in two ways: We assume that
the BC that result in the large length-scale limit (i) do not break the $O(n)$ symmetry and (ii) are associated with the respective most stable renormalization-group (RG) fixed point. For parallel orientation this simply means that Dirichlet BC $\bm{\phi}=\bm0$ hold asymptotically \cite{DGR03,DRG03,DR04,Die05}. In the case of perpendicular orientation, \emph{two} BC hold on either boundary plane. These simplify in the large-length-scale limit to the conditions that both the order parameter $\bm{\phi}$ and its normal derivative $\partial_n\bm{\phi}$ \cite{Die05,DSP06} vanish.

In the next section we  introduce the models and specify their Hamiltonians including the boundary terms they involve in the cases of free boundary planes. In section~\ref{sec:BC} we first give the mesoscopic BC that result from the boundary contributions to the classical equations of motion in the cases of parallel and perpendicular slab orientations. Assuming that the values of the surface interaction constants comply with the above-mentioned conditions (i) and (ii), we then recapitulate which asymptotic boundary conditions are encountered in the limit of large length scales. In section~\ref{sec:bgptrg} we recall the background on the renormalization of these models at and below their upper critical dimensions $d^*(m)=4+m/2$ and the field-theoretic RG approach to bulk and surface critical behavior at  LP required for our subsequent analysis of the Casimir forces.  We then turn to the calculation of fluctuation-induced forces at the LP. The case of parallel slab orientation is dealt with in section~\ref{sec:calcpar}, that of perpendicular  orientation in section~\ref{sec:CAperp}. Section~\ref{sec:concl} provides a brief summary and concluding remarks. Finally, there are 3 appendixes describing technical details.

\section{Models}\label{sec:Mod}

\begin{figure}[htb]
\centering
\includegraphics[width=0.6\textwidth]{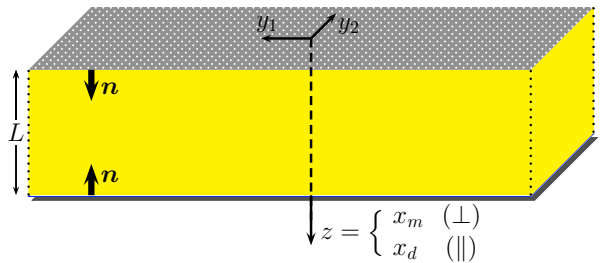}
\caption{Slab geometries considered: For perpendicular and parallel orientations of the surface planes at $z=0$ and $z=L$, the $z$-axis is along the $\alpha$-direction $x_m$ and the $\beta$-direction $x_d$, respectively. The remaining $\bm{y}$-directions are $\alpha$- or $\beta$-directions. Further, $\bm{n}$ is the inward-pointing normal.}\label{fig:geometry}
\end{figure}

The models we consider involve an $n$-component order-parameter field
$\bm{\phi}(\bm x)= (\phi_a(\bm x),
a=1,\ldots,n)$ defined on the slab
$\mathfrak{V}=\mathbb{R}^{d-1}\times [0,L]$ of $d$-dimensional space
$\mathbb{R}^d$. We write position vectors as
$\bm x =(\bm{y},z)$, where
$\bm{y}\in\mathbb{R}^{d-1}$ and $z\in[0,L]$ are the coordinates alongside and across the slab, respectively (see \fref{fig:geometry}). Parallel orientation means that $z$ is a $\beta$-coordinate, perpendicular that it is an $\alpha$-coordinate. Without loss of generality, we can take the first $m$ Euclidean axes as $\alpha$-directions.  We choose $z$ to be the $\beta$- or $\alpha$-coordinate with the largest index, so that $z=x_d$ and  $z=x_m$ for parallel and perpendicular orientations of the boundary planes, respectively.
 
The slab is assumed to have infinite area $A=\infty$ of its boundary planes. We can therefore choose PBC along the corresponding $d-1$ principal $y$-directions for convenience. Depending on whether we take PBC or
FBC in the $z$-direction, the slab $\mathfrak{V}$ has no boundary, $\partial\mathfrak{V}=\emptyset$, or its boundary consists of the
two ($d-1$)-dimensional confining hyperplanes $\mathfrak{B}_1$ at
$z=0$ and $\mathfrak{B}_2$ at $z=L$. In the latter case, we orient the
boundary such that the normal $\bm{n}$ on
$\partial\mathfrak{V}=\mathfrak{B}\equiv\mathfrak{B}_1\cup\mathfrak{B}_2$
points into the interior of $\mathfrak{V}$.

Ignoring any long-range interactions, we choose the Hamiltonians to be
of the form
\begin{equation}
  \label{eq:Hamfb}
  {\mathcal H}={\int_{\mathfrak{V}}}{\mathcal L}_{\rm b}(\bm x)\,
  {\rm d} V+ \delta_{{\rm BC},{\rm free}}\sum_{j=1}^2
  {\int_{\mathfrak{B}_j}}{\mathcal L}^{\|,\perp}_j(\bm x)
  \,{{\rm d}}A\;, 
\end{equation}
where ${\mathcal L}_{\rm b}(\bm x)$ and ${\mathcal
  L}^{\|,\perp}_j(\bm x)$ are functions of the field
$\bm{\phi}(\bm x)$ and its spatial first and second
derivatives
$\partial_\gamma\bm{\phi}(\bm x)
\equiv\partial\bm{\phi(\bm x)}/\partial
x_\gamma$ and
$\partial_\gamma\partial_{\gamma'}\bm{\phi}(\bm x)$.
 Following \cite{DS00a,SD01}, we
choose the bulk density appropriate for the description of
the (multi)critical behavior at
$m$-axial Lifshitz points as
\begin{eqnarray}
  \label{eq:Lb}\fl
 {\mathcal L}_{\rm b}(\bm x)&=&\frac{\mathring{\sigma}}{2}
  \Big(\sum_{\alpha=1}^m\partial_\alpha^2\bm{\phi} \Big)^2
 +\frac{\mathring{\rho}}{2}
\sum_{\alpha=1}^m{(\partial_\alpha\bm{\phi})}^2
+\frac{1}{2}\sum_{\beta=m+1}^d{(\partial_\beta\bm{\phi})}^2
+\frac{\mathring{\tau}}{2}\bm{\phi}^2
+\frac{\mathring{u}}{4!}\,|\bm{\phi}|^4\,.
\end{eqnarray}
On the level of Landau theory, the LP is located at $\mathring\tau=\mathring\rho=0$.

When $m>1$, this choice of ${\mathcal L}_{\rm b}$ ignores potential anisotropies breaking the rotational invariance in the $m$-dimensional Euclidean $\alpha$-space $\mathbb{R}^m$, such as a term proportional to the hypercubic invariant $\sum_{\alpha=1}^m(\partial^2_\alpha\bm{\phi})^2$. According to the two-loop calculation reported in  \cite{DSZ03}, such a contribution is relevant in the RG sense. We omit it here, as well as similar less symmetric terms, for the sake of simplicity. In the uniaxial case $m=1$, this is no restriction.

Throughout the paper, we shall consider PBC and FBC. In the former case,
\begin{equation}
  \bm{\phi}(\bm{y},z+L)
  =\bm{\phi}(\bm{y},z)
\end{equation}
and the boundary terms in equation~\eqref{eq:Hamfb} are absent, as indicated.

In the case of FBC, different boundary densities
$\mathcal{L}_j^{\|}(\bm x)$ and
$\mathcal{L}_j^{\perp}(\bm x)$ (dictated by
relevance/irrelevance considerations) must be chosen to define
appropriate minimal models for slabs with parallel or perpendicular
orientation. Work on semi-infinite systems%
\footnote{For general background on the field-theoretic RG analysis of
  critical behavior in semi-infinite systems, see
  \cite{Die86a} and \cite{Die97}.}
\cite{DGR03,DRG03,DR04,Die05,DSP06} suggests the choices
\begin{equation}
  \label{eq:L1par}
  {\mathcal L}_j^\|(\bm x)=\frac{1}{2}\,\mathring{c}_j\phi^2
  +\frac{\mathring{\lambda}_j}{2}\,
  \sum_{\alpha=1}^m(\partial_\alpha\bm{\phi})^2
\end{equation}
and
\begin{eqnarray}
  \label{eq:L1perp}
  {\mathcal L}^\perp_j(\bm x)&=&\frac{1}{2}\,\mathring{c}_j^\perp\phi^2
+\mathring{b}_j\bm{\phi}\partial_n\bm{\phi}
+\frac{1}{2}\,\mathring{\lambda}_j^\perp
(\partial_n\bm{\phi})^2
\nonumber\\ &&\strut
  +\sum_{\alpha=1}^{m-1}\left[ \frac{1}{2}\,\mathring{\lambda}_j^{\parallel}\,
  (\partial_\alpha\bm{\phi})^2 +\mathring{f}_j\,
  (\partial_\alpha\bm{\phi})
  \partial_n\partial_\alpha\bm{\phi}\right],
\end{eqnarray}
where we temporarily allow  all bare interaction constants
$\mathring{c}_j,\ldots,\mathring{f}_j$ to take different values on the
two boundary planes $\mathfrak{B}_j$, $j=1,2$.

\section{Mesoscopic and asymptotic boundary conditions}\label{sec:BC}

As discussed elsewhere \cite{Die86a,Die97}, for actions of the
form~\eqref{eq:Hamfb}, the boundary contributions
to the classical equations of motion  give us ``mesoscopic BC'' that hold in an
operator sense. We call them mesoscopic because they hold on the length scale beyond which the chosen continuum description is valid. They must be distinguished from the asymptotic large-length-scale BC one encounters  at criticality. Since the boundary contributions to the action differ in the cases of parallel and perpendicular orientations, so do the mesoscopic BC. We first consider the case of parallel slab orientation.

\subsection{Parallel slab orientation }\label{sec:BCpar}
In this case, the action is defined by equations~\eqref{eq:Hamfb},
\eqref{eq:Lb} and \eqref{eq:L1par}. The mesoscopic BC become \cite{DGR03,DRG03}
\begin{equation}
  \label{eq:bcpar}
  \partial_n\bm{\phi}(\bm x)
  =\bigg(\mathring{c}_j-\mathring{\lambda}_j
\sum_{\alpha=1}^m
\partial^2_\alpha\bigg)\bm{\phi}(\bm x)\;,\quad
\bm x \in\mathfrak{B}_j\,.
\end{equation}

The bare interaction constants $\mathring{c}_j$ and $\mathring{\lambda}_j$ have engineering dimension
$[\mathring{c}_j]=\mu$ and $[\mathring{\lambda}_j]=\mu^0$, where $\mu$ is a momentum scale. The former are
relevant at Gaussian fixed points with $\mathring{c}_j=\mathring{\lambda}_j=0$. Beyond Landau
theory, the associated renormalized quantities $c_j$, which are
proportional to the bare deviations
$\delta\mathring{c}_j=\mathring{c}_j-\mathring{c}_{\rm sp}$ from a
cutoff-dependent special value $\mathring{c}_{\rm sp}$,
take over the roles of relevant surface scaling
fields \cite{DR04}.%
  \footnote{When deviations $\rho\propto
  \mathring{\rho}-\mathring{\rho}_{\rm LP}$ of the bare bulk variable
  $\mathring{\rho}$ from its value $\mathring{\rho}_{\rm LP}$ at the
  bulk LP are considered, the corresponding linear surface
  scaling fields $g_{c_j}$ actually are linear combinations of $c_j$
  and $\rho$, the renormalized counterpart of $\mathring{\rho}$.
  Details, which will not be needed below, can be found in
  \cite{DR04}.}   We assume that the initial values of the
renormalized surface variables $c_j$ and $\lambda_j$ lie in the basin of attraction of the fixed point with $c_j=\infty$ describing the so-called ordinary surface transitions of the semi-infinite systems with one surface plane at $\mathfrak{B}_j$ and the respective other at $z=\infty$ or $z=-\infty$. As is known from \cite{DGR03},
\cite{DRG03} and \cite{DR04}, Dirichlet BC hold at
this fixed point at both boundary planes $\mathfrak{B}_j$. To study
the corresponding asymptotic behavior, one can set the bare values
$\mathring{c}_j=\infty$. Then the regularized bare theory satisfies
Dirichlet BC. The values of the bare variables
$\mathring{\lambda}_j\ge 0$ do not matter. To investigate the critical
Casimir forces in this case of parallel slab orientation and large-scale
Dirichlet BC, we can simply set
$\mathring{c}_1=\mathring{c}_2=\infty$ and drop the boundary
contributions $\propto \mathring{\lambda}_j$. Equivalently, we can impose
Dirichlet BC
\begin{equation}\label{eq:bcaspar}
  \bm{\phi}(\bm x)=\bm0\;,
  \quad\forall \bm x \in\mathfrak{B}_1\cup\mathfrak{B}_2,
\end{equation}
on the regularized bare theory and drop
the boundary terms $\int_{\mathfrak{B}_j}\mathcal{L}^{\|}_j\,\rmd{A}$ of the action.

\subsection{Perpendicular slab orientation}

In this case, the action is defined by equations~\eqref{eq:Hamfb},
\eqref{eq:Lb}, and \eqref{eq:L1perp}. Since $z$ now is an $\alpha$-direction, the classical equation of motion is of fourth rather than of second order in $\partial_z$. In conformity with this we find two (instead of one) mesoscopic 
BC at either boundary plane, namely  \cite{Die05,DSP06}
\begin{equation}
  \label{eq:bcperp1}
  {\Bigg\{\mathring{\sigma}\partial_n^3+(\mathring{b}_j-
  \mathring{\rho})\partial_n+\mathring{c}_j^\perp
  -[\mathring{\lambda}_j^{\|}+(\mathring{f}_j-\mathring{\sigma})\partial_n]
  \sum_{\alpha=1}^{m-1}\partial_\alpha^2\Bigg\}}\bm{\phi}(\bm{x}) =\bm{0}
\end{equation}
and
\begin{equation}
  \label{eq:bcperp2}
\Bigg[-\mathring{\sigma}\,\partial_n^2
+\mathring{\lambda}_j^{\perp}\partial_n 
  +\mathring{b}_j-(\mathring{f}_j+\mathring{\sigma})
  \sum_{\alpha=1}^{m-1}\partial_\alpha^2\Bigg]
  \bm{\phi}(\bm x)=\bm{0}\;\;\mbox{ for }\bm{x}\in\mathfrak{B}_j.
\end{equation}

The problem of boundary critical behavior at Lifshitz points has been
investigated to a much lesser degree for the case of perpendicular
orientation of boundary planes. However, as shown in
\cite{Die05} and \cite{DSP06}, one can benefit from
analogous simplifications when analyzing the asymptotic behavior at
the corresponding so-called ``ordinary'' surface transition (described by the most stable fixed point with bulk LP behavior). Let us recapitulate the
essence of the argument leading to the conclusion that 
\begin{equation}
  \label{eq:bcasperp}
  \bm{\phi}(\bm x)=\bm{0},\quad
  \partial_n\bm{\phi}(\bm x)=\bm{0},
  \quad\forall \bm x \in\mathfrak{B}_1\cup\mathfrak{B}_2,
\end{equation}
are the appropriate large-length-scale BC to consider. Noting that
$\partial_n$ now has engineering dimension $\mu^{1/2}$, one finds that
the interaction constants of the boundary densities
$\mathcal{L}^{\perp}_j$ with the largest momentum dimensions are
$\mathring{c}^{\perp}_j\sim\mu^{3/2}$ and $\mathring{b}_j \sim\mu$.
Thus both give rise to scaling fields that are relevant at the Gaussian fixed points where they vanish. For generic positive initial values, they will be driven to $\infty$ under RG transformations $\mu\to\mu\ell$ in the infrared (IR) limit $\ell\to 0$. We therefore
take the limits $\mathring{c}^{\perp}_j\to\infty$ and $\mathring{b}_j\to\infty$ at the outset. Upon dividing the boundary conditions~\eqref{eq:bcperp1} and \eqref{eq:bcperp2} by
$\mathring{c}^{\perp}_j$ and $\mathring b_j$, respectively, and
scaling all interaction constants according to their $\mu$ dimensions,
one arrives at the asymptotic conditions~\eqref{eq:bcasperp} for the
regularized bare theory.

The upshot of these  considerations and  those in the previous subsection is the following. To investigate the asymptotic form of the Casimir forces at the LP in the case where the surface interaction constants are subcritically enhanced at both boundary planes, we can impose the Dirichlet BC~\eqref{eq:bcaspar}  and the BC~\eqref{eq:bcasperp}, depending on whether the slab orientation is parallel or perpendicular, and omit the boundary terms $\sum_j\int_{\mathfrak{B}_j}\mathcal{L}_j^{\parallel,\perp}\,\rmd A$ of the respective actions~\eqref{eq:Hamfb}.

\section{Background and renormalization group }\label{sec:bgptrg}

We wish to investigate the residual free energy $f_{\mathrm{res}}$ in $d=d^*(m)-\epsilon$ dimensions by means of the RG approach. Our aims are to confirm the asymptotic behavior~\eqref{eq:fres} and compute the Casimir amplitudes $\Delta_{\|,\perp}^{\mathrm{BC}}$ for parallel and perpendicular slab orientation under the mentioned BC  using RG-improved perturbation theory. To this end we must first supply some background on free propagators and the renormalization of the models. We begin by recalling a number of required results of the free bulk theory.

\subsection {Free bulk propagators}\label{sec:fbp}

For notational convenience we decompose the position vector
$\bm{x}=(\bm{y},z)$ into its $m$-dimensional $\alpha$-component $\bm{r}=(x_\alpha)$ and $(d-m)$-dimensional $\beta$-component  $\bm{s}=(x_\beta)$, writing $\bm{x}= (\bm{r}, \bm{s})$. Here and below  it is understood that $\alpha$  runs from $1$ to $m$, while $\beta=m+1,\dots,d$. Likewise, we use the notation $\bm{q}=(\bm{k},\bm{p})$ with $\bm{k}=(q_\alpha)$ and $\bm{p}=(q_\beta)$ for $d$-dimensional wave vectors $\bm{q}$ conjugate to $\bm{x}$ and their $\alpha$- and $\beta$-components. 

The free bulk propagator $G^{(d,m)}_{\mathrm{b}}$ of the disordered phase in $d$ dimensions follows from the Gaussian part of the bulk density $\mathcal{L}_{\mathrm{b}}$ given in equation~\eqref{eq:Lb}. For general non-negative values of $\rhob$ and $\tb$, it can be written as
\begin{equation}
 \label{eq:Gb}
 G^{(d,m)}_{\mathrm{b}}(\bm{x}|\sigb;\tb,\rhob)=\int_{\bm{q}=(\bm{k},\bm{p})}^{(d)}\frac{\rme^{\rmi (\bm{r}\cdot\bm{k}+\bm{s}\cdot\bm{p})}}{p^2+\sigb k^4+\rhob k^2+\tb}\,,
 \end{equation}
 where 
\begin{equation}
 \label{eq:intab}
\quad\int_{\bm{q}}^{(d)}\equiv\int_{\mathbb{R}^d}\frac{\rmd^dq}{(2\pi)^d}=\int_{\bm{k}}^{(m)}\int_{\bm{p}}^{(d-m)}
\end{equation}
denotes a conveniently normalized momentum integral. By rotational invariance in the $\bm{r}$- and $\bm{s}$-subspaces, the position dependence of $G^{(d,m)}_{\mathrm{b}}(\bm{x}|\sigb;\tb,\rhob)$ reduces to a dependence on the lengths $r=|\bm{r}|$ and $s=|\bm{s}|$ of the $\alpha$- and $\beta$-components of $\bm{x}$.

We shall use dimensional regularization. Unless stated otherwise, all momentum integrations will therefore not be restricted by a large-momentum cutoff $\Lambda$. Being interested in properties at the LP (such as the Casimir amplitudes $\Delta^{\mathrm{BC}}$), we will normally set $\rhob=\tb=0$ and work with the free LP propagator $G^{(d,m)}_{\mathrm{b}}(\bm{x})\equiv G^{(d,m)}_{\mathrm{b}}(\bm{x}|\sigb;0,0)$. This can be written as \cite{DS00a,SD01}
\begin{equation}
 \label{eq:GbLP}
 G^{(d,m)}_{\mathrm{b}}(\bm{x})=\sigb^{-m/4}s^{-2+\epsilon}\,\Phi_{m,d}(\upsilon)\,,\quad\upsilon\equiv\sigb^{-1/4}rs^{-1/2}
\end{equation}
with the scaling function
\begin{equation}
 \label{eq:Phiint}
 \Phi_{m,d}(\upsilon)=\int_{\bm{q}}^{(d)}\frac{\rme^{\rmi(\bm{\upsilon}\cdot\bm{k}+\hat{\bm{s}}\cdot\bm{p})}}{p^2+k^4}\;,
\end{equation}
where $\hat{\bm{s}}$ is an arbitrarily oriented unit vector in $\mathbb{R}^{d-m}$. For general values of $m$ and $d$, the latter function is a difference of generalized hypergeometric functions ${}_1F_2$ (equal to a Fox-Wright $\Psi$ function \cite{SD01}). Its explicit form can be found, for example, in equation~(A2) of \cite{DS00a} or equations~(11) and (13) of \cite{SD01}.  For our purposes here, it will be sufficient to know its Taylor expansion
\begin{equation}\label{eq:Phitaylor}
\Phi_{m,d}(\upsilon)=2^{-2-m}\pi^{(2\epsilon-6-m)/4}\sum_{l=0}^\infty\frac{(-1)^l}{2^{2l}l!}\,\frac{\Gamma[(2+l-\epsilon)/2]}{
\Gamma[(2+m+2l)/2)}\,
\,\upsilon^{2l}
\end{equation}
and its asymptotic expansion for large $\upsilon$,
\begin{equation}
 \label{eq:Phiasexp}\fl
\Phi_{m,d}(\upsilon)\mathop{\approx}_{\upsilon\to\infty} 2^{3-m-2\epsilon}\pi^{(2\epsilon-6-m)/4}\,\upsilon^{2\epsilon-4}\sum_{l= 0}^\infty\frac{(-1)^l}{ l!}\,\frac{2^{4l}\,\Gamma(2-\epsilon+2l)}{\Gamma[(m-2+2\epsilon-4l)/4]}
\,\upsilon^{-4l}.
\end{equation}

From the former one can infer the value at $\upsilon=0$:
\begin{equation}
 \label{eq:Phi0}
 \Phi_{m,d}(0)=2^{-2-m}\pi^{(2\epsilon-6-m)/4}\, \frac{\Gamma(1-\epsilon/2)}{\Gamma[(2+m)/ 4]}\,.
\end{equation}
We shall also need the leading asymptotic behavior of $\Phi_{m,d}$ for large $\upsilon$ below. According to equation~\eqref{eq:Phiasexp}, we have
\begin{equation}
 \label{eq:Phiasinfty}
 \Phi_{m,d}(\upsilon)\mathop{=}_{\upsilon\to\infty}\Phi_{m,d}^{(\infty)}\,\upsilon^{2\epsilon-4}+O\big(\upsilon^{2\epsilon-8}\big)
\end{equation}
with
\begin{equation}
 \label{eq:Phiinfty}
 \Phi_{m,d}^{(\infty)}=2^{3-m-2\epsilon}\pi^{(2\epsilon-6-m)/4}\,\frac{\Gamma(2-\epsilon)}{\Gamma[(m-2+2\epsilon)/4]}\,.
 \end{equation}
Note that the coefficient $ \Phi_{m,d}^{(\infty)}$ vanishes on the whole line $d=3+m$. This reflects the fact that the function $\Phi_{m,3+m}(\upsilon)$ decays $\sim\rme^{-\upsilon^2/4}$ (cf.\ equation~(19) of \cite{SD01}). 

\subsection{Reparametrizations and renormalization group}\label{sec:REP}

Let us introduce the (dimensionally regularized) $N$-point cumulants (connected correlation functions)
\begin{equation}
 \label{sec:GNdef}
 G_{L;a_1,\ldots,a_N}^{(N)}(\bm{x}_1,\ldots,\bm{x}_N)=\langle\phi_{a_1}(\bm{x}_1)\cdots\phi_{a_N}(\bm{x}_N)\rangle^{\mathrm{cum}}
\end{equation}
and denote their bulk (infinite-space) analogues as $G_{\mathrm{b};a_1,\ldots,a_N}^{(N)}(\bm{x}_1,\ldots,\bm{x}_N)$. The renormalization of the latter functions in $d=d^*(m)-\epsilon\le d^*(m)$ dimensions has been explained in detail in \cite{DS00a,SD01,DGR03}. Their ultraviolet (UV) divergences induced by the singularity of the free bulk propagator $G^{(d,m)}_{\mathrm{b}}(\bm{x})$ at $\bm{x}=\bm{0}$ can be absorbed by means of the reparametrizations
\begin{eqnarray}\label{eq:bulkrep}
  \bm{\phi}=Z_\phi^{1/2}\,\bm{\phi}_{\mathrm{R}}\;,\quad
  \mathring{\sigma}=Z_\sigma\,\sigma\;,
\quad  \mathring{u}\,{\mathring{\sigma}}^{-m/4}\,F_{m,\epsilon}=
   \mu^\epsilon\,Z_u\,u\;,\nonumber
\\
\mathring{\tau}-\mathring{\tau}_{\mathrm{LP}}=
\mu^2\,Z_\tau{\big(\tau+A_\tau\,\rho^2\big)}\;,
\quad
\left(\mathring{\rho}-\mathring{\rho}_{\mathrm{LP}}\right)\,
{\mathring{\sigma}}^{-1/2}=\mu\,Z_\rho\,\rho\,,
\end{eqnarray}
where $\mu $ is a momentum scale. Following \cite{SD01} and \cite{DGR03}, we choose the factor  $F_{m,\epsilon }$ that is absorbed in the renormalized coupling constant as
\begin{equation}
  \label{eq:Fmeps}
F_{m,\epsilon}=
\frac{\Gamma{\left(1+{\epsilon/ 2}\right)}
\,\Gamma^2{\left(1-{\epsilon/ 2}\right)}\,
\Gamma{\left({m}/{4}\right)}}{(4\,\pi)^{({8+m-2\,\epsilon})/{4}}\,
\Gamma(2-\epsilon)\,\Gamma{\left({m}/{2}\right)}}\,.
\end{equation}
The LP is located at $(\tb,\rhob)=(\tb_{\mathrm{LP}},\rhob_{\mathrm{LP}})$. In a theory regularized by means of a large-momentum cutoff $\Lambda$, the renormalization functions $\tb_{\mathrm{LP}}$ and $\rhob_{\mathrm{LP}}$  would diverge $\sim\Lambda^2$ and $\sim \Lambda$, respectively. In our perturbative approach based on dimensional regularization, they vanish. Results to order $u^2$ for the renormalization factors $Z_\phi$, $Z_\sigma$, $Z_\rho$, $Z_\tau$ and $Z_u$ may be found in equations~ (40)--(50) of \cite{SD01}. The function $A_\tau$ is given to $\Or(u)$ in equation~(17) of \cite{DGR03}. 

Upon varying $\mu$ at fixed values of the bare variables $\mathring{u}$, $\sigb$,   $\rhob$, and $\tb$, we find that the renormalized bulk functions $G^{(N)}_{\mathrm{b;R}}=Z_\phi^{-N/2}G^{(N)}_{\mathrm{b}}$ satisfy the RG equations
\begin{equation}
\label{eq:RGE}
\left(\mathcal{D}_\mu+\frac{N}{2} \,\eta_\phi\right)G^{(N)}_{\mathrm{b};a_1,\ldots,a_N;\mathrm{R}}(\bm{x}_1,\ldots,\bm{x}_N)=0
\end{equation}
with
\begin{equation}
\label{eq:Dmu}
\mathcal{D}_\mu=\mu\,\partial_\mu+\sum_{g=u,\sigma,\rho,\tau}\beta_g\partial_g\,.
\end{equation}
The beta functions $\beta_g$  are defined by
\begin{equation}
\label{eq:betagdef}
\beta_g\equiv\mu\,\partial_\mu|_0\,g\,,\;\;g=u,\sigma,\rho,\tau,
\end{equation}
where $\partial_\mu|_0$ means a derivative at fixed bare interactions constants. They can be expressed in terms of the exponent functions
\begin{equation}
\label{eq:etas}
\eta_g(u)=\mu\,\partial_\mu |_0\ln{Z}_g\,,\;g=\phi,u,\sigma,\rho,\tau,
\end{equation}
and
\begin{equation}
\label{eq:btau}
b_\tau(u)=A_\tau[\mu\,\partial_\mu |_0\ln A_\tau+\eta_\tau-2\eta_\rho]
\end{equation}
as
\begin{equation}
\label{eq:betags}
\beta_g=
\cases{
-u[\epsilon+\eta_u(u)],&$g=u$,\\
-\sigma\eta_\sigma(u),&$g=\sigma$,\\
-\rho[1+\eta_\rho(u)],&$g=\rho$,\\
-\tau[2+\eta_\tau(u)]-b_\tau(u)\rho^2,&$g=\tau$.
}
\end{equation}

The reparametrizations~\eqref{eq:bulkrep}  also suffice to absorb the UV singularities of the $N$-point functions $G_{L;a_1,\ldots,a_N}^{(N)}$ for films of finite thickness $L$ under PBC, irrespective of whether the orientation is parallel or perpendicular. This can be concluded from the form of the free film propagator (see, e.g., \cite[chapter IV]{Die86a}),
\begin{equation}\label{eq:GLIM}
G_{L;\|,\perp}^\mathrm{PBC}(\bm x-\bm x')=\sum_{j=-\infty}^\infty
G^{(d,m)}_{\mathrm{b}}(\bm x-\bm x'+jL\hat{\bm{z}})\,,
\end{equation} 
where $\hat{\bm{z}}$ is a unit vector along the $z$-axis.
The crux of the argument is that all primitive UV singularities must be induced by the $j=0$ summand (the bulk term), because the remaining $j\ne 0$ summands are finite at $\bm{x}=\bm{0}$. Accordingly, the RG equations~\eqref{eq:RGE} hold also for the film cumulants $G^{(N)}_{L;a_1,\dots,a_N}$ under PBC.

In the case of FBC one generally expects additional primitive UV singularities localized at the boundaries. Whenever they occur, counterterms with support on the boundary planes $\mathfrak{B}_j$ (``surface counterterms'') must be added to the action. However, we can benefit from the simplifying features of the chosen Dirichlet BC~\eqref{eq:bcaspar} and \eqref{eq:bcasperp} for parallel and perpendicular orientations. These BC ensure that the renormalized correlation functions $G_{L;\{a_i\};\mathrm{R}}^{(N)}=Z^{-N/2}_\phi\,G_{L;\{a_i\}}^{(N)}$ defined via the reparametrizations~\eqref{eq:bulkrep} become UV finite even when some (or all) fields $\phi_{a_i}(\bm{x}_i)$ are located on the boundary. Surface counterterms would be required in the case of parallel slab orientation if we wanted to renormalize cumulants involving normal derivatives $\partial_n\phi_a$ in addition to fields $\phi_{a}(\bm{x})$ with $\bm{x}\in\mathfrak{V}$ \cite{DGR03,DRG03}. For perpendicular surface orientation,  such cumulants would still be UV finite owing to the stronger BC that both $\bm{\phi}$ and $\partial_n\bm{\phi}$ vanish. On the other hand, cumulants involving the boundary operators $\partial^2_n\bm{\phi}$ would require surface counterterms \cite{DSP06}.

In order to investigate the residual free energy $f_{\mathrm{res}}(L;\ldots)$ by means of the RG approach, we must know  how to renormalize this quantity. The bulk and surface excess free energy densities  $f_{\mathrm{b}}$ and $f_{\mathrm{s}}$ are known to require additional additive counterterms. The subtractions these counterterms provide do not depend on $L$. Thus, they cancel in the difference of the free-energy densities defining $f_{\mathrm{res}}(L;\ldots)$. Using the reparametrizations~\eqref{eq:bulkrep} to express bare variables in terms of renormalized ones therefore gives us UV finite renormalized residual free-energy densities 
\begin{equation}
 \label{eq:fresR}
f_{\mathrm{res;R}}(L;\tau,\rho,u,\sigma,\mu)=f_{\mathrm{res}}(L;\tb,\rhob,\mathring{u},\mathring{\sigma}),
\end{equation}
both for PBC as well as  either one of the  Dirichlet BC~\eqref{eq:bcaspar} and 
\eqref{eq:bcasperp}. 

As an immediate consequence we obtain the RG equation
\begin{equation}
\label{eq:RGEfres}
\mathcal{D}_\mu f_{\mathrm{res;R}}(L;\tau,\rho,u,\sigma,\mu)=0\,.
\end{equation}
Solving it at the LP $\tau=\rho=0$ via characteristics and using dimensional analysis gives
\begin{eqnarray}
\label{eq:fresas}
\fl f_{\mathrm{res;R}}(L;0,0,u,\sigma,\mu)=f_{\mathrm{res;R}}[L;0,0,\bar{u}(\ell),\bar{\sigma}(\ell),\mu\ell] 
\nonumber\\ \fl
=\cases{(\mu\ell)^{d-m-1}\left(\frac{\mu^2\ell^2}{\bar{\sigma}(\ell)}\right)^{m/4}\,f_{\mathrm{res;R}}[L\mu\ell;0,0,\bar{u}(\ell),1,1]\,,&$\varsigma=\|$,\\
(\mu\ell)^{d-m}\left(\frac{\mu^2\ell^2}{\bar{\sigma}(\ell)}\right)^{(m-1)/4}\,f_{\mathrm{res;R}}[L(\mu^2\ell^2/\bar{\sigma}(\ell))^{1/4};0,0,\bar{u}(\ell),1,1]\,,&$\varsigma=\perp$,
}
\end{eqnarray}
where the variable $\varsigma$ specifies the orientation. Further, $\bar{u}(\ell)$ and $\bar{\sigma}(\ell)$ are examples of running variables $\bar{g}(\ell)$ defined in the usual way as solutions to the flow equations
\begin{equation}
 \label{eq:gbar}
\ell\frac{\rmd}{\rmd\ell}\,\bar{g}(\ell)=\beta_g[\bar{\tau}(\ell),\bar{\rho}(\ell),\bar{u}(\ell),\bar{\sigma}(\ell)],\quad g=\tau,\ldots,\sigma,
\end{equation}
subject to the initial conditions
\begin{equation}
 \label{eq:initg}
\bar{g}(\ell=1)=g.
\end{equation}

In the large-length-scale limit $\ell\to 0$, the running coupling constant approaches the IR-stable root $u^*$,  whose $\epsilon$ expansion 
\begin{equation}
 \label{eq:ustar}
u^*=\frac{6\epsilon}{n+8}+\Or(\epsilon^2)
\end{equation}
is known to second order \cite{SD01} but will be needed only to first order in the following. 

The variable $\bar{\sigma}(\ell)$ behaves as
\begin{equation}
 \label{eq:sigmabaras}
\bar{\sigma}(\ell)\mathop{\approx}_{\ell\to 0} E_\sigma^*(u)\,\ell^{-\eta_\sigma(u^*)}\,\sigma,
\end{equation}
where the fixed-point value of $\eta_\sigma$ is related to the anisotropy exponent $\theta$ by
\begin{equation}
\label{eq:theta} 
\theta=[2+\eta_\sigma(u^*)]/4
\end{equation}
and $E_\sigma^*(u)\equiv E_\sigma(u^*,u)$ means the value of the trajectory integral
\begin{eqnarray}
\label{eq:Esigma}
E_\sigma(\bar{u},u)&=&\exp\left\{\int_u^{\bar{u}}\rmd u'\,\frac{\eta_\sigma(u^*)-\eta_\sigma(u')}{\beta_u(u')}\right\}\nonumber\\ &=&\exp\left(\int_0^\ell \frac{\rmd\ell '}{\ell '}\left\{4\theta -2-\eta_\sigma[\bar{u}(\ell)]\right\}\right)
\end{eqnarray}
at $\bar{u}=u^*$.

We now choose $\ell$ such that the scaled $L$-dependent arguments of $f_{\mathrm{res;R}}$ in equation~\eqref{eq:fresas} become $1$, obtaining
\begin{eqnarray}
 \label{eq:fresRresult}\fl
f_{\mathrm{res;R}}^{\mathrm{BC}}(L;0,0,u,\sigma,\mu)\mathop{\approx}_{L\to \infty}\cases{\mu^{m(1-2\theta)/2}\,(E_\sigma^*\sigma)^{-m/4}\,\Delta^{\mathrm{BC}}_\| L^{-\zeta_\|},&$\varsigma=\|$,\nonumber\\
\mu^{(d-m)(2\theta-1)/(2\theta)}\,(E_\sigma^*\sigma)^{(d-m)/(4\theta)}\,\Delta^{\mathrm{BC}}_\perp L^{-\zeta_\perp},&$\varsigma=\perp$,
}\\
\end{eqnarray}
with
\begin{equation}
\label{eq:Deltafres}
\Delta^{\mathrm{BC}}_{\varsigma}=f^{\mathrm{BC}}_{\mathrm{res;R}}(1;0,0,u^*,1,1),
\end{equation}
where $\zeta_{\|,\perp}$ are the decay exponents introduced in equation~\eqref{eq:zetas}. 
$\mathrm{BC}$ stands for either PBC  or else the asymptotic ones associated with the ordinary surface transitions, at both boundary planes. Depending on the orientation, the latter are given by the Dirichlet BC \eqref{eq:bcaspar} ($\varsigma=\|$) or equation~\eqref{eq:bcasperp}  ($\varsigma=\perp$). In either case,  we use the notation $\mathrm{BC}=(\mathrm{O,O})$ to refer to these boundary conditions.

\subsection{Casimir amplitudes as universal ratios}\label{sec:Delunivratios}
Note that the result~\eqref{eq:fresRresult} involves, besides the momentum scale $\mu$ and the variable $\sigma$ (both of which are there for dimensional reasons), the nonuniversal amplitude $E_\sigma^*(u)$. On the other hand, it is not difficult to see that the amplitudes $\Delta^{\mathrm{BC}}_\varsigma$ can be written as universal ratios of nonuniversal quantities. To this end, let us introduce the second moment finite-size correlation lengths
\begin{equation}
 \label{eq:xialpha}
\xi_{L;\alpha,\varsigma}^{2}=\frac{1}{m}\sum_{\alpha=1}^m\overline{(x_\alpha-x'_\alpha)^2}
\end{equation}
and
\begin{equation}
 \label{eq:xibeta}
\xi_{L;\beta,\varsigma}^{2}=\frac{1}{d-m}\sum_{\beta=m+1}^d\overline{(x_\beta-x'_\beta)^2},
 \end{equation}
 where the averages $\overline{h(\bm{x},\bm{x}')}$ are defined by
 \begin{equation}
 \label{eq:overline}
\overline{h(\bm{x},\bm{x}')}=\frac{\int_{\mathfrak{V}}\rmd^dx\int_{\mathfrak{V}}\rmd^dx'\,h(\bm{x},\bm{x}')\,G^{(2)}_{L;\mathrm{R}}(\bm{x},\bm{x}')}{\int_{\mathfrak{V}}\rmd^dx\int_{\mathfrak{V}}\rmd^dx'\,G^{(2)}_{L;\mathrm{R}}(\bm{x},\bm{x}')}.
\end{equation}

 As indicated, these 	quantities depend on the orientation $\varsigma=\|$ or $\varsigma=\perp$ whenever $L<\infty$. In the limit $L\to\infty$ they approach the corresponding (orientation-independent) second-moment bulk correlation lengths $\xi_{\mathrm{b};\alpha}\equiv \xi_{\infty;\alpha,\varsigma}$ and $\xi_{\mathrm{b};\beta}\equiv \xi_{\infty;\beta,\varsigma}$. Note that we assumed  the bulk density~\eqref{eq:Lb} to be isotropic in both the $\alpha$- and $\beta$-subspaces. Therefore, the bulk correlation lengths can equivalently be written as bulk moments $\xi_{\mathrm{b};\alpha}^{2}=\big[\overline{x_\alpha^2}\big]_{\mathrm{b}}$ and $\xi_{\mathrm{b};\beta}^{2}=\big[\overline{x_\beta^2}\big]_{\mathrm{b}}$ for arbitrary choices of $\alpha=1,\dots,m$ and $\beta=m+1,\ldots,d$, where $\big[\overline{h(\bm{x})}\big]_{\mathrm{b}}$ denotes 
 the bulk average $\int_{\mathbb{R}^d}\rmd^dx\,h(\bm{x})\, G^{(2)}_{\mathrm{b,R}}(\bm{x})/\int_{\mathbb{R}^d}\rmd^dx\, G^{(2)}_{\mathrm{b,R}}(\bm{x})$.
  A violation of the bulk density's isotropy in $\beta$-space would correspond to what is usually called ``weak anisotropy''. It means that the bulk moments $\big[\overline{x_\beta^2}\big]_{\mathrm{b}}$ associated with different values of $\beta$  have the same power-law singularity $\sim\tau^{-\nu_{L2}}$ as the LP is approached along a path with $\rho=0$ but may have different (nonuniversal) amplitudes. Getting rid of such a weak anisotropy can be achieved in a straightforward fashion by transforming to order parameter densities that yield a squared gradient term in $\beta$-space of the form assumed in equation~\eqref{eq:Lb}. To achieve this, one can proceed along lines analogous to those taken in \cite{Doh08,DC09} to cope with weak anisotropy at critical points. 
  
  Anisotropies in $\alpha$-space are more dangerous. The two-loop RG analysis \cite{DSZ03} of isotropy-violating linear combinations of $(\partial^2\bm{\phi}/\partial x_\alpha^2)^2$ to the bulk density $\mathcal{L}_{\mathrm{b}}$ revealed that the fixed point in $d^*(m)-\epsilon$ dimensions is unstable to such  perturbations. For this reason we explicitly rule out the presence of such isotropy-breaking terms.

Now, the RG equations $\mathcal{D}_\mu\,\xi_{L;\alpha,\varsigma}=\mathcal{D}_\mu\,\xi_{L;\beta,\varsigma}=0$ in conjunction with dimensional analysis yield
\begin{eqnarray}
 \label{eq:xialphaRGsol}
\xi_{L;\alpha,\varsigma}(u,\tau,\rho,\sigma,\mu)&=&(\mu\ell)^{-1/2}\,\bar{\sigma}^{1/4}\,\xi_{\bar{L};\alpha,\varsigma}(\bar{u},\bar{\tau},\bar{\rho},1,1)\nonumber\\&\mathop{\approx}_{\ell\to 0}&\mu^{-1/2}\left[E_\sigma^*(u)\sigma\right]^{1/4}\ell^{-\theta}\,\xi_{\bar{L};\alpha,\varsigma}(u^*,\bar{\tau},\bar{\rho},1,1)
\end{eqnarray}
and
\begin{eqnarray}
 \label{eq:xibetaRGsol}
 \xi_{L;\beta,\varsigma}(u,\tau,\rho,\sigma,\mu)&=&(\mu\ell)^{-1}\xi_{\bar{L};\beta,\varsigma}(\bar{u},\bar{\tau},\bar{\rho},1,1)\nonumber\\ &
 \mathop{\approx}_{\ell\to 0}&(\mu\ell)^{-1}\xi_{\bar{L};\beta,\varsigma}(u^*,\bar{\tau},\bar{\rho},1,1),
 \end{eqnarray}
 where 
 \begin{equation}
 \label{eq:Lbar}\fl
\bar{L}=\cases{
\mu \ell L&for $\varsigma=\|$,\\
\mu^{1/2}\ell^{1/2}{\bar{\sigma}}^{-1/4}L\mathop{\approx}_{\ell\to 0} \mu^{1/2}[E^*_\sigma(u)\, \sigma]^{-1/4}\ell^{\theta}L&for $\varsigma=\perp$.}
\end{equation}

Setting $\tau=\rho=0$ and choosing the flow parameter $\ell=\ell_{\varsigma}$ again such that $\bar{L}=1$ gives 
\begin{equation}\fl
\label{eq:xialphaampl}
\xi^{\mathrm{BC}}_{L;\alpha,\varsigma}(u,0,0,\sigma,\mu)\mathop{\approx}_{L\to\infty}\cases{\Xi^{\mathrm{BC}}_{\alpha,\|}\,L^\theta= \mu^{\theta-1/2}\,(E_\sigma^*\sigma)^{1/4} \xi_{1;\alpha,\|}^{*\mathrm{BC}}\,L^{\theta}&for $\varsigma=\|$,\\
\Xi_{\alpha,\perp}^{\mathrm{BC}}\, L\,= \xi_{1;\alpha,\perp}^{*\mathrm{BC}}\,L&for $\varsigma=\perp$,}
\end{equation}
and
\begin{equation}\fl
\label{eq:xibetaampl}
\xi^{\mathrm{BC}}_{L;\beta,\varsigma}(u,0,0,\sigma,\mu)\mathop{\approx}_{L\to\infty}\cases{\Xi^{\mathrm{BC}}_{\beta,\|}\,L= \xi_{1;\beta,\|}^{*\mathrm{BC}}\,L&for $\varsigma=\|$,\\
\Xi_{\beta,\perp}^{\mathrm{BC}}\, L^{1/\theta}= \mu^{\frac{1-2\theta}{2\theta}} (E_\sigma^*\sigma)^{-\frac{1}{4\theta}} \,\xi_{1;\beta,\perp}^{*\mathrm{BC}}\,L^{1/\theta}&for $\varsigma=\perp$,}
\end{equation}
where we have introduced the universal numbers $\xi_{1;\gamma,\varsigma}^{*\mathrm{BC}} \equiv \xi_{1;\gamma,\varsigma}^{\mathrm{BC}}(u^{*},0,0,1,1)$ with $\gamma=\alpha, \beta$ and $\varsigma= \|, \perp$. Owing to the presumed isotropy of our model in both the $\alpha$- and $\beta$-subspaces, the amplitudes $\Xi_{\alpha,\varsigma}^{\mathrm{BC}}$ are the same for all possible $m$ choices of $\alpha$. Likewise, $\Xi_{\beta,\varsigma}^{\mathrm{BC}}$ is independent of the choice of $\beta$.

Using the above results and definitions, one easily  concludes that the Casimir amplitudes can be written as
\begin{eqnarray}
 \label{eq:Deltaparratio}\fl
\Delta^{\mathrm{BC}}_\varsigma\nonumber 
=\lim_{L\to\infty}f^{\mathrm{BC}}_{\mathrm{res;R}}(L;0,0,u,\sigma,\mu)\\
\times\left[\frac{\xi^{\mathrm{BC}}_{L;\alpha,\varsigma}(u,0,0,\sigma,\mu)}{\xi_{1;\alpha,\varsigma}^{*\mathrm{BC}}}\right]^{m-\delta_{\varsigma,\perp}}\,\left[\frac{\xi^{\mathrm{BC}}_{L;\beta}(u,0,0,\sigma,\mu)}{\xi_{1;\beta,\varsigma}^{*\mathrm{BC}}}\right]^{d-m-\delta_{\varsigma,\|}},
\end{eqnarray}
i.e., the dependences on $\mu$, $\sigma$ and the nonuniversal scale factor $E_\sigma^*$ drop out of these ratios. Note that this definition would even work if weak ``diagonal'' anisotropies were allowed in both the $\alpha$- and $\beta$-subspace. By such anisotropies we mean the kind that can be transformed away by simple rescalings of the Cartesian coordinates $x_\alpha$ and $x_\beta$, respectively.  In other words, we assume that the metrics in the $\alpha$- and $\beta$-subspaces provided  by the respective gradient square terms in the Hamiltonian remain diagonal \cite{DC09}. The case of lattice systems that lack hypercubic symmetries in the $\alpha$- and $\beta$-subspaces, such as systems with monoclinic lattices, requires somewhat more thought since the metric in full space must be diagonalized. To this end, the considerations made in \cite{Doh08,DC09} for weakly anisotropic $\phi^4$ models near critical points (CP) must be appropriately adapted and generalized.

\section {Calculation of fluctuation-induced interactions at the bulk LP}
 \label{sec:paror}
Having established the RG predictions for the asymptotic behavior of the residual free energy density, we now turn to the calculation of the Casimir amplitudes $\Delta^{\mathrm{BC}}_{\varsigma}$ by means of RG-improved perturbation theory in $d=d^*(m)-\epsilon$ dimensions. We first consider the case of parallel slab orientation. It is computationally less involved  than that of perpendicular film orientation. Furthermore, one can  proceed in close analogy to the previous analysis of Casimir forces at critical points \cite{KD91,KD92a,DGS06,GD08}.

\subsection{Parallel slab orientation} \label{sec:calcpar}

We begin by considering the exactly solvable Gaussian case with $\mathring{u}=0$. In this noninteracting case, we have $\sigb=\sigma$, $\tb=\mu^2 \tau$, and $\rhob=\mu\rho$. To obtain the Gaussian analogue of the result for $f_{\mathrm{res}}$, we must insert the classical value $1/2$ for the anisotropy exponent $\theta$ both in equation~\eqref{eq:fresRresult} and expression~\eqref{eq:zetas} for the decay exponent $\zeta_\|$. In addition, the nonuniversal amplitude $E_\sigma^*$ must be dropped. This yields
\begin{equation}
\label{eq:fresparGauss}
f^{\mathrm{BC}}_{\mathrm{res}}(L;0,0,0,\sigma,\mu)=\sigma^{-m/4}\Delta_{\|,\mathrm{G}}^{\mathrm{BC}}\,L^{-(d-1-m/2)}.
\end{equation}
In ~\ref{app:fg} we show that the Casimir amplitudes $\Delta_{\|,\mathrm{G}}^{\mathrm{BC}}$ are given by
\begin{eqnarray}
\label{eq:DeltaparGauss}
\Delta_{\|,\mathrm{G}}^{\mathrm{PBC}}(d,m,n)&=&2^{d-m/2}\Delta_{\|,\mathrm{G}}^{(\mathrm{O,O})}(d,m,n) \nonumber \\
&=&-n\,C_m\,\frac{\Gamma[(d-m/2)/2]\,\zeta(d-m/2)}{\pi^{(d-m/2)/2}}
\end{eqnarray}
with
\begin{equation}
 \label{eq:Cm}
C_m=\frac{\pi^{(2-m)/4}}{2^{m}\,\Gamma[(m+2)/4]}.
\end{equation}

The factor $C_m$ is a ratio of geometric factors one encounters in the calculation of Feynman integrals  for bulk critical behavior at LP and critical points (CP) that involve the respective free bulk propagators $G^{(d,m)}_{\mathrm{b}}(\bm{0}|\sigb;0,0)$ and $G^{(d,0)}_{\mathrm{b}}(\bm{0}|\sigb;0,0)$ at coinciding positions $\bm{x}'=\bm{x}$ (tadpole graphs).  Two examples of its occurrence are the ratios
\begin{equation}
 \label{eq:Cmorig}
C_m=\frac{F_{m,\epsilon}}{F_{0,\epsilon}}=\frac{\Phi_{m,4+m/2-\epsilon}(0)}{\Phi_{0,4-\epsilon}(0)},
\end{equation}
where the $\Phi_{m,d}(0)$ are the values of the functions~\eqref{eq:Phiint} at $\upsilon=0$.

To understand the origin of $C_m$, note that $G^{(d,m)}_{\mathrm{b}}(\bm{0}|\sigb;0,0)$ according to equation~\eqref{eq:Gb} is given by a  momentum integral of the form $\int_{\bm{p}}^{(d-m)}\int_{\bm{k}}^{(m)}h(p^2+\sigb k^4)$ with $h(q^2)=1/q^2$.   Such integrals can be expressed in terms of the ($d-m/2$)-dimensional momentum integrals $\int_{\bm{q}}^{(d-m/2)}h(q^2)$: using hyperspherical coordinates in  $\bm{k}$ and $\bm{p}$~space, one can transform to the variable $K=\sigb^{1/2}\,k^2$ and then introduce the polar coordinates $p=q\cos\theta$ and $K=q\sin\theta$, following \cite{ML78}.  The integration measure $p^{d-m-1}\,k^{m-1}\,\rmd{p}\,\rmd{k}$ of the radial integrations becomes $\sigb^{-m/4}\,(q\cos\theta)^{d-m-1}(q\sin\theta)^{(m-2)/2}\,q\,\rmd{q}\,\rmd{\theta}/2$, and a straightforward calculation leads to
\begin{equation}\label{eq:Cmrel}
\int^{(m)}_{\bm{k}}\int_{\bm{p}}^{(d-m)}h(p^2+\sigb k^4)=\sigb^{-m/4}\,C_m\int_{\bm{q}}^{(d-m/2)}h(q^2).
\end{equation}

Returning to the above results~\eqref{eq:fresparGauss} and  \eqref{eq:DeltaparGauss}, let us note that they must reduce to their known CP analogues in the limit $m\to 0$. The Casimir amplitudes $\Delta_{\mathrm{CP,G}}^{\mathrm{(O,O)}}(d,n)$  of the free field theory with action density $\mathcal{L}_b=\sum_{\gamma=1}^n(\partial_\gamma\bm{\phi})^2/2$ can be read off from  \cite[equation~(4.4)]{Sym81} and  are explicitly given in \cite[equation~(5.6)]{KD92a}. Their analogues $\Delta_{\mathrm{CP,G}}^{\mathrm{PRB}}(d,n)$ can be found in \cite[equation~(5.7), 3rd line]{KD92a}.%
\footnote{Note that a minus sign is missing in this equation of \cite{KD92a}. This is an evident misprint.}
Taking into account that $C_0=1$,  one sees that the corresponding expressions are indeed recovered from our equations~\eqref{eq:fresparGauss} and  \eqref{eq:DeltaparGauss} in the limit $m\to 0$. 

A straightforward consequence  is that the Casimir amplitudes~\eqref{eq:DeltaparGauss} can be expressed in terms of their CP analogues as
\begin{equation}
\label{eq:CPLPrel}
\Delta_{\|,\mathrm{G}}^{\mathrm{BC}}(d,m,n)=C_m\,\Delta_{\mathrm{CP,G}}^{\mathrm{BC}}(d-m/2,n).
\end{equation}
In fact, the analogue of this relation between $\Delta_{\|}^{\mathrm{BC}}(d,m,n)$ and $\Delta_{\mathrm{CP}}^{\mathrm{BC}}(d-m/2,n)$ turns out to hold in the limit $n\to\infty$ for $2+m/2<d\le 4+m/2$. To make this statement precise, let us define the large-$n$ amplitudes
\begin{equation}
 \label{eq:DeltaSMPBC}
\Delta_{\|,\infty}^{\mathrm{BC}}(d,m)=\lim_{n\to\infty}\Delta_{\|}^{\mathrm{BC}}(d,m,n)/n\quad\mbox{for }\mathrm{BC}=\mathrm{PBC},\,(\mathrm{O,O}),
\end{equation}
and their CP analogues $\Delta_{\mathrm{CP},\infty}^{\mathrm{BC}}(D)$, where we assume that $2<D=d-m/2\le 4$,  i.e.\ that $D$ lies between the lower and upper critical bulk dimensions $d_*(m=0)=2$ and $d^*(m=0)=4$, respectively. Our claim is that these amplitudes obey the relation
\begin{equation}
 \label{eq:DeltaparrelSM}
\Delta_{\|,\infty}^{\mathrm{BC}}(d,m)=C_m\,\Delta_{\mathrm{CP},\infty}^{\mathrm{BC}}(d-m/2), \quad\mathrm{BC}=\mathrm{PBC},\,(\mathrm{O,O}).
\end{equation}
Its derivation is given in \ref{app:largen}. For PBC, this is straightforward. The case of $(\mathrm{O,O})$ BC requires more thought because the loss of translation invariance perpendicular to the boundary planes entails that the exact solution in the limit $n\to\infty$  involves an effective $z$-dependent pair interaction, which must be determined self-consistently \cite{Kno73,BM77,Comtesse09}.

One may wonder whether analogues of the relations~\eqref{eq:CPLPrel} and \eqref{eq:DeltaparrelSM} might hold in general. We emphasize that there is no reason to expect this. The origin of their validity in the free-field case and  the large-$n$ limit is that the corresponding amplitudes involve only  Feynman integrals of a sort that equation~\eqref{eq:Cmrel} can be applied.%
\footnote{For example, the free energy of the Gaussian theory involves the function $h(q^2)=\ln q^2$.}
However, at higher orders of the loop expansion Feynman graphs involving powers of free propagators between different points in position space occur. Relation~\eqref{eq:Cmrel} is not applicable to them.  As we shall see below, the relations~\eqref{eq:CPLPrel} and \eqref{eq:DeltaparrelSM} carry over to the small-$\epsilon$ expansion to the order of our calculation. The reason is the same as before: only Feynman graphs involving free propagators at  coinciding points contribute. At order $\epsilon^2$, three-loop graphs involving powers of free propagators between distinct positions contribute. These must be expected to invalidate the analogues of relations~\eqref{eq:CPLPrel} and \eqref{eq:DeltaparrelSM}.

We next consider the general interacting case $\mathring{u}\ne 0$ with finite $n$,  addressing the issue of the series expansions of the Casimir amplitudes $\Delta_\|^{(\mathrm{O,O})}(d,m,n)$ and $\Delta_\|^{\mathrm{PBC}}(d,m,n)$ in $\epsilon=d^*(m)-d$. We assert that these quantities have the Taylor  and fractional power series expansions
\begin{equation}
 \label{eq:DeltaparDD}\fl
\frac{\Delta_\|^{(\mathrm{O,O})}}{n}=\frac{-C_m\pi^2}{1440}\Bigg\{1+\Bigg[\frac{\gamma_E-1}{2}+\ln(2\sqrt{\pi})- \frac{\zeta'(4)}{\zeta(4)}-\frac{5(n+2)}{4(n+8)}\Bigg]\epsilon\Bigg\}+\Or(\epsilon^2)
 \end{equation}
 and
 \begin{eqnarray}
 \label{eq:DeltaparPBC}
 \frac{\Delta_\|^{\mathrm{PBC}}}{n}&=&C_m\Bigg\{\frac{-\pi^2}{90}+\frac{\pi^2 \epsilon}{180}\left[1-\gamma_E-\ln \pi+\frac{2\,\zeta'(4)}{\zeta(4)}+\frac{5(n+2)}{2(n+8)}\right]\nonumber\\ &&\strut -\frac{\pi^2}{9\sqrt{6}}\bigg(\frac{n+2}{n+8}\bigg)^{3/2}\,\epsilon^{3/2}\Bigg\}+\mathrm{o}(\epsilon^{3/2}),
\end{eqnarray}
respectively. Here $\gamma_E=-\Gamma'(1)=0.577215\ldots$ is the Euler-Mascheroni constant,  and $\zeta(s)$ means the Riemann zeta function. 

As is borne out by the second equation, the small-$\epsilon$ expansion of $\Delta_\|^{\mathrm{PBC}}$ also involves half-integer powers of $\epsilon$. Moreover, beyond the given order $\epsilon^{3/2}$, additional half-integer powers of $\epsilon$ appear together with powers of $\ln \epsilon$. For example, there is a term  proportional to $\epsilon^{5/2} \ln\epsilon$. This is completely analogous to what was found in the CP case for $\Delta_\|^{\mathrm{PBC}}$ \cite{DGS06,GD08,DG09}. The existence of such terms should become clear as we outline the derivation of the expansion~\eqref{eq:DeltaparPBC} below. Before we turn to this matter, a few other remarks are appropriate.

Relations~\eqref{eq:CPLPrel} and \eqref{eq:DeltaparrelSM} suggest to compare the above results with the analogous series expansions of  $\Delta^{(\mathrm{O,O})}_{\mathrm{CP}}$ and $\Delta^{\mathrm{PBC}}_{\mathrm{CP}}$ in $\epsilon=4-d$. The comparison shows that all explicitly displayed orders in equations~\eqref{eq:DeltaparDD} and \eqref{eq:DeltaparPBC} agree with their counterparts for $\Delta^{(\mathrm{O,O})}_{\mathrm{CP}}$  and $\Delta^{\mathrm{PBC}}_{\mathrm{CP}}$ obtained in \cite{KD91,KD92a} and \cite{DGS06,GD08}, respectively, up to the factor $C_m$ and the fact that $\epsilon$ means $4-d$ in the CP case rather than $4+m/2-d$. 

An immediate important consequence of this finding and equation~\eqref{eq:DeltaparrelSM} should be mentioned. It was shown elsewhere \cite{GD08} that the large-$n$ limit of the fractional $\epsilon=4-d$ expansion of $\Delta^{\mathrm{PBC}}_{\mathrm{CP}}$ to $\Or(\epsilon^{3/2})$ is in conformity with the $\epsilon=4-d$ expansion of $\Delta^{\mathrm{PBC}}_{\mathrm{CP},\infty}$. Thus the large-$n$ limit of the series expansion \eqref{eq:DeltaparPBC} in $\epsilon=4+m/2-d$ to $\Or(\epsilon^{3/2})$ is also consistent with the small-$\epsilon$ expansion of the large-$n$ amplitude $\Delta_{\|,\infty}^{\mathrm{BC}}(d,m)$ to this order in $\epsilon$. 

Using relation~\eqref{eq:DeltaparrelSM} for PBC, we can determine $ \Delta_{\|,\infty}^{\mathrm{PBC}}(3,1)$ in a straightforward manner by computing $ \Delta_{\mathrm{CP},\infty}^{\mathrm{PBC}}(5/2)$. According to \cite{DDG06} and \cite{GD08}, $\Delta_{\mathrm{CP},\infty}^{\mathrm{PBC}}(d)$ is given by
\begin{equation}
\Delta_{\mathrm{CP},\infty}^{\mathrm{PBC}}(d)=-\frac{A_d}{d}\,\mathcal{R}_d^{d/2}-\frac{4\pi\,Q_{d+2,2}(\mathcal{R}_d)}{\mathcal{R}_d}.
\end{equation}
Here $A_d$ and the function $Q_{d,2}(y)$ are defined by
\begin{equation}
\label{eq:Ad}
A_{d}=-(4\pi)^{-d/2}\Gamma(1-d/2),
\end{equation}
and
\begin{eqnarray}\label{eq:Qd2def}
Q_{d,2}(y)&=&\frac{y}{2}\left[\sum_{k\in 2\pi\mathbb{Z}}\int_{\bm{p}}^{(d-1)}-\int_{\bm{q}=(\bm{p},k)}^{(d)}\right]\frac{1}{q^2+y}\nonumber\\
&=&y^{d/2}(2\pi)^{-d/2}\sum_{j=1}^\infty\left(j\sqrt{y}\right)^{-(d-2)/2}\,K_{(d-2)/2}(j\sqrt{y}),
\end{eqnarray}
where $K_{\nu}(z)$ is the Macdonald function (modified Bessel function of the second kind). Further,  $\mathcal{R}_d$ is the solution to
\begin{equation}
A_d\,\mathcal{R}_d^{(d-2)/2}=2\,\mathcal{R}_d^{-1}\,Q_{d,2}(\mathcal{R}_d).
\end{equation}

Solving the latter equation at $d=5/2$ by numerical iteration yields
\begin{equation}\mathcal{R}_{5/2}=0.5358\ldots,
\end{equation}
from which the numerical values of the Casimir amplitudes 
\begin{equation}\label{eq:CPlargenpar}
\Delta_{\mathrm{CP},\infty}^{\mathrm{PBC}}(5/2)= -0.2337\ldots
\end{equation}
and
\begin{equation}\label{eq:largenpar}
\Delta_{\|,\infty}^{\mathrm{PBC}}(3,1)=-0.1269\ldots
\end{equation}
follow.

It will suffice to give a brief outline of how the above small-$\epsilon$ expansion results~\eqref{eq:DeltaparDD} and \eqref{eq:DeltaparPBC} were obtained. In order to determine  $\Delta_\|^{(\mathrm{O,O})}$ to $\Or(\epsilon)$, one can start from the two-loop expression for the total free energy $F$, set $\tb=\tb_{\mathrm{LP}}$ and $\rhob=\rhob_{\mathrm{LP}}$, subtract the bulk and surface contributions to $F(k_BTA)^{-1}$ to identify the residual free energy density $f_{\mathrm{res}}(L;\ldots)$, express the latter in terms of renormalized variables $u$ and $\sigma$, and then evaluate $f_{\mathrm{res};R}$ at the fixed-point value $u^*$ given in equation~\eqref{eq:ustar} and set $L=\sigma=\mu=1$. If we proceeded in this manner in the case of PBC, we would encounter a zero-mode problem: The Fourier transform of the free propagator $G^{\mathrm{PBC}}_{L;\|}$ at zero momentum $(\bm{k},\bm{p})=\bm{0}$ becomes massless at the LP and hence IR singular. To see this, note that the spectral decompositions of the free propagators $G_{L;\|}^{\mathrm{BC}}$ at $\tb=\rhob=0$ for the two BC in question read
\begin{equation}
\label{eq:Gparspecdec}\fl
G_{L;\|}^{\mathrm{BC}}(\bm{x},\bm{x}')=\int_{\bm{k}}^{(m)}\int_{\bm{p}}^{(d-m-1)}\sum_r
\frac{\varphi^{\mathrm{BC}}_r(z)\,[\varphi^{\mathrm{BC}}_r(z')]^*}{
p^2+P_r^2+\sigb k^4}\,\exp\left[\rmi \sum_{\gamma=1}^{d-1}q_\gamma(x_\gamma-x'_\gamma)\right],
\end{equation}
where 
\begin{equation}
\label{eq:varphiBC}\fl
\varphi^{\mathrm{BC}}_r(z)=\frac{1}{\sqrt{L}}
\left\{
\begin{array}{llll}
  \exp(\rmi P_rz),& P_r=2\pi r/L,  &r\in\mathbb{Z},&\mathrm{BC}=\mathrm{PBC},   \\[\medskipamount]
\sqrt{2}\sin(P_rz),  & P_r=r\pi/L,  &r=1,\ldots,\infty,&\mathrm{BC}=(\mathrm{O,O}),  \end{array}
\right.
\end{equation}
are the normalized eigenfunctions  of the operator $-\partial_z^2$ on the interval $[0,L]$, $P_r^2$ are its eigenvalues, and the sums $\sum_r$ run over the indicated BC-dependent sets.

For PBC, the lowest eigenvalue $P_r^2=P_0^2$ vanishes. Hence Landau theory  predicts the film to have a LP at the same location $(\tb,\rhob)=(0,0)$ as the bulk. This is a deficiency of Landau theory. For $(\mathrm{O,O})$  BC (corresponding to Dirichlet BC), this does not happen. Since the lowest eigenvalue $P^2_1$ is strictly positive, none of the modes becomes critical at the bulk LP.

To cope with this zero-mode problem, we proceed as in \cite{DGS06}, \cite{GD08} and \cite{DG09}. We split off the ($P_r=0$) zero-mode contribution 
\begin{equation}
 \label{eq:varphi}
\bm{\varphi}(\bm{y})= \delta_{\mathrm{BC},\mathrm{PBC}}\,
L^{-1/2}\int_0^L \rmd z\,\bm{\phi}(\bm{y},z)
\end{equation}
of the order parameter in the case of PBC, writing
\begin{equation}
\label{eq:splitphi}
\bm{\phi}(\bm{y},z)=
L^{-1/2}\,\bm{\varphi}(\bm{y})+\bm{\psi}(\bm{y},z).
\end{equation}
Depending on the BC, one either has
\begin{equation}
\label{eq:psipropDD}
\bm{\psi}(\bm{x})\equiv\bm{\phi}(\bm{x})\mbox{ and }
\bm{\varphi}(\bm{y})\equiv\bm{0},
\;\;\mathrm{BC}=(\mathrm{O,O}),
\label{eq:PBC}
\end{equation}
or
\begin{equation}
\int_0^L\rmd{z}\,\bm{\psi}(\bm{y},z)=0,\;\;\mathrm{BC}=\mathrm{PBC}.
\end{equation}

The field $\bm{\psi}$ gives a contribution $F_\psi$, defined by
\begin{equation}
\label{eq:Fpsi}
\frac{F_{\psi}}{k_BT}=-\ln\Tr_{\psi}\rme^{-\mathcal{H}[\bm{\psi}]},
\end{equation}
to the total free energy. We denote the associated reduced free-energy density as
\begin{equation}
\label{eq:fpsi}
f_\psi(L)\equiv\lim_{A\to\infty}\frac{F_{\psi}}{k_BTA}.
\end{equation}
Its Feynman graph expansion to two-loop order is given by
\begin{equation}\label{eq:fpsigraphs}
-f_{\psi}(L)A=\,\,\raisebox{-8pt}{\includegraphics[width=22pt,clip]{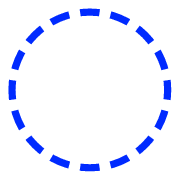}}\,\,+\,\,\raisebox{-14.75pt}{\includegraphics[width=18pt,clip]{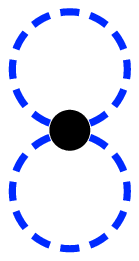}}\,\,+\Or(3\mbox{-loops}),
\end{equation}
where the dashed blue line represents the free $\psi$-propagator, i.e., the free propagator $G^{\mathrm{BC}}_{\psi,L;\|}$ one obtains from the spectral decomposition~\eqref{eq:Gparspecdec} of $G^{\mathrm{BC}}_{L;\|}$ by omitting the ($P_0\,{=}\,0$)-mode contribution in the case of $\mathrm{PBC}$.  Noting that the zero-mode contribution can be written in terms of free bulk propagator in $d-1$ dimensions, one concludes that
\begin{eqnarray}
 \label{eq:Gpsi}\fl
G^{\mathrm{BC}}_{\psi,L;\|}(\bm{x};\bm{x}')&=&\int_{\bm{k}}^{(m)}\int_{\bm{p}}^{(d-m-1)}\sum_{P_r\ne 0}
\frac{\varphi^{\mathrm{BC}}_r(z)\,[\varphi^{\mathrm{BC}}_r(z')]^*}{
p^2+P_r^2+\sigb k^4}\,\exp\left[\rmi \sum_{\gamma=1}^{d-1}q_\gamma(x_\gamma-x'_\gamma)\right]
\nonumber\\ &=&G^{\mathrm{BC}}_{L;\|}(\bm{x};\bm{x}')-\delta_{\mathrm{BC},\mathrm{PBC}}\,L^{-1}\,G^{(d-1,m)}_{\mathrm{b}}(\bm{y}-\bm{y}').
\end{eqnarray}
The Feynman graphs displayed in equation~\eqref{eq:fpsigraphs} are computed in \ref{app:fg}.

Since $f_\psi(L)$ does not involve a zero mode, it is perturbatively well defined at the bulk LP. Its value at $u=u^*$ has an expansion in non-negative integer powers of $\epsilon$. For $\mathrm{BC}=(\mathrm{O,O})$, $f_\psi$ is identical to the total reduced free-energy area density $f(L)$, whose definition should be clear by analogy with equation~\eqref{eq:fpsi}. However, for $\mathrm{PBC}$ the contribution associated with $\bm{\varphi}$ must be added. We therefore write
\begin{equation}
 \label{eq:fLfpsifvarphi}
f(L)=f_{\psi}(L)+f_{\varphi}(L)\,\delta_{\mathrm{BC},\mathrm{PBC}}.
\end{equation}
To determine $f_{\varphi}(L)$, we substitute the decomposition~\eqref{eq:splitphi} of $\bm{\phi}$ into the Hamiltonian and define an effective $(d-1)$-dimensional action $\mathcal{H}_{\mathrm{eff}}[\bm{\varphi}]$ by tracing out $\bm{\psi}$. This yields
\begin{equation}
\label{eq:Heff}
\rme^{-\mathcal{H}_{\mathrm{eff}}[\bm{\varphi}]}=\rme^{F_\psi/k_BT}\,\Tr_\psi\rme^{-\mathcal{H}[\bm{\phi}[\bm{\psi},\bm{\varphi}]]}
\end{equation}
with
\begin{equation}
\label{eq:Hamsplit}
\mathcal{H}[\bm{\phi}[\bm{\psi},\bm{\varphi}]]=\mathcal{H}[\bm{\psi}]+\mathcal{H}[L^{-1/2}\bm{\varphi}]+ \mathcal{H}_{\mathrm{int}}[\bm{\varphi},\bm{\psi}],
\end{equation}
where the contribution depending exclusively on $\bm{\varphi}$ and the interaction part are given by
\begin{eqnarray}
\label{eq:Hvarphipart}
 \mathcal{H}[L^{-1/2}\bm{\varphi}]&=&\int\rmd^{d-1}y\Bigg[
 \frac{\mathring{\sigma}}{2}
  \Big(\sum_{\alpha=1}^m\partial_\alpha^2\bm{\varphi} \Big)^2
+\frac{1}{2}\sum_{\beta=m+1}^{d-1}{(\partial_\beta\bm{\varphi})}^2
+\frac{\mathring{u}}{4!L}\,|\bm{\varphi}|^4\nonumber\\ &&\strut 
+\frac{\mathring{\rho}}{2}
\sum_{\alpha=1}^m{(\partial_\alpha\bm{\varphi})}^2+\frac{\mathring{\tau}}{2}\bm{\varphi}^2
 \Bigg]
 \end{eqnarray}
 and
\begin{equation}
 \label{eq:Hint}
  \mathcal{H}_{\mathrm{int}}[\bm{\varphi},\bm{\psi}]=\frac{\mathring u}{12L}\int_\mathfrak{V}\rmd{V}
\left[\varphi^2\psi^2+2(\bm{\varphi}\cdot\bm{\psi})^2+
2\sqrt{L}\,(\bm{\varphi}\cdot\bm{\psi})\psi^2\right],
 \end{equation}
 respectively.

The effective action can be written in terms of an average $\langle \dots\rangle_\psi$ with the path probability density $\rme^{F_\psi-\mathcal{H}[\bm{\psi}]}\mathcal{D}[\bm{\psi}]$ as
\begin{equation}
\label{eq:Heffpsiav}
\mathcal{H}_{\mathrm{eff}}[\bm{\varphi}]=\mathcal{H}[L^{-1/2}\bm{\varphi}]-\ln\left\langle\rme^{-\mathcal{H}_{\mathrm{int}}[\bm{\varphi},\bm{\psi}]}\right\rangle_\psi.
\end{equation}
In terms of the latter, the contribution $f_\varphi(L)$ becomes
\begin{equation}
\label{eq:Fvarphi}
f_\varphi(L)=\lim_{A\to\infty} \frac{F_{\varphi}}{k_BTA}=-\lim_{A\to\infty}\frac{1}{A}\ln\Tr_{\varphi}\rme^{-\mathcal{H}_{\mathrm{eff}}[\bm{\varphi}]}.
\end{equation}
Since we ultimately wish to use RG-improved perturbation theory, we need some results of the loop expansion of $ \mathcal{H}_{\mathrm{eff}}[\bm{\varphi}]$:
\begin{equation}
 \label{eq:Heffle}
 \mathcal{H}_{\mathrm{eff}}[\bm{\varphi}]=\mathcal{H}[L^{-1/2}\bm{\varphi}]+ \mathcal{H}^{[1]}_{\mathrm{eff}}[\bm{\varphi}]+\Or(2\mbox{-loops}).
\end{equation}
The one-loop contribution is given by
\begin{eqnarray}
\label{eq:Heff1}
\mathcal{H}^{[1]}_{\mathrm{eff}}[\bm{\varphi}]&=&\frac{1}{2}\Tr \ln\left[\bm{1}+\frac{\mathring{u}}{6L}\,G^{\mathrm{PBC}}_{\psi,L;\|}\left(\delta_{\alpha\beta}\varphi^2+2\varphi_\alpha\varphi_\beta\right)\right]\nonumber\\
&=&\raisebox{-0.6em}{\includegraphics[width=35pt]{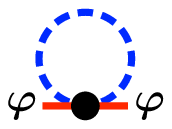}}\;\; +\;\raisebox{-1.05em}{\includegraphics[width=85pt]{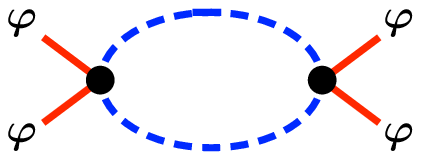}}\,+\ldots.
\end{eqnarray}
Up to the order of our approximation ($\epsilon^{3/2}$) only the first graph explicitly shown in equation~\eqref{eq:Heff1} contributes. It produces a shift $\tb\to\tb+ \delta\mathring{\tau}_{L,\|}^{\mathrm{PBC}}$ of the coefficient $\tb$ in $\mathcal{H}[L^{-1/2}\bm{\varphi}]$ whose evaluation at $\tb=\rhob=0$ and $d=4+m/2-\epsilon$ is straightforward, giving
\begin{equation}
  \label{eq:shiftper}
   \delta\mathring{\tau}_{L,\|}^{\mathrm{PBC}}
=\sigb^{-m/4} C_m\,\mathring{u}\,\frac{n+2}{6}\,
\frac{\Gamma(1-\epsilon/2)\, \zeta(2-\epsilon)}{2\pi^{2-\epsilon/2}\,
  L^{2-\epsilon}}.
\end{equation}
The comparison with equation~(14) of \cite{DGS06} shows that this shift is related to its CP analogue $\delta\mathring{\tau}_{L,\mathrm{CP}}^{\mathrm{PBC}}$ in the expected manner:
\begin{equation}
\label{eq:shiftLPCP}
 \delta\mathring{\tau}_{L,\|}^{\mathrm{PBC}}(d,m)=\sigb^{-m/4} C_m\,\delta\mathring{\tau}_{L,\mathrm{CP}}^{\mathrm{PBC}}(d-m/2).
\end{equation}

Both $\Or(\mathring{u})$ shifts are proportional to $\zeta(2-\epsilon)$, where $\zeta(s)$ is the Riemann zeta function. The latter is known to have a unique analytic continuation to the entire complex plane, excluding the point $s=1$, at which it has a simple pole with residue $1$ \cite[chapter 23]{AS72}. This pole corresponds to a logarithmic IR singularity: When $d=3+m/2$ ($\epsilon=1$), the free bulk propagator $G_{\mathrm{b}}^{(d,m)}(\bm{x})$
decays at the LP for large $|\bm{x}|$ so slowly that $G_{L;\|}^{\mathrm{PBC}}(\bm{x},\bm{x})$ diverges logarithmically. Thus it acquires an IR pole, just as the shifts~\eqref{eq:shiftper} do. On the other hand, these shifts have an expansion  about $\epsilon=0$ (i.e., about $d_*(m)=4+m/2$  and $d^*(0)=4$, respectively). 
Upon expressing $\delta\mathring{\tau}_{L,\|}^{\mathrm{PBC}}$ in terms of renormalized variables and  using $\zeta(2-\epsilon)=\pi^2/6+\Or(\epsilon)$, one arrives at the expansion 
\begin{equation}
\label{eq:deltatauren}
\delta\mathring{\tau}_{L,\|}^{\mathrm{PBC}}=\frac{u}{L^2}\Bigg[\frac{\pi^2(n+2)}{9}+\Or(u,\epsilon)\Bigg].
\end{equation}

Note that this quantity is positive when evaluated at the IR-stable root $u^*$ of $\beta_u$. This is just as in the CP case: owing to this $L$-dependent shift provided by the coupling to the nonzero modes, the $\bm{\varphi}$ field does not become critical at the LP. Therefore, we can use the free  propagator of the Hamiltonian~\eqref{eq:Hvarphipart} with $\rhob=0$ and  $\delta\mathring{\tau}_{L,\|}^{\mathrm{PBC}}$ substituted for $\mathring{\tau}$ in the Feynman graph expansion of $f_\varphi^{\mathrm{PBC}}(L)$. This propagator is nothing else than the bulk propagator $G_{\mathrm{b}}^{(d-1,m)}(\bm{y}|\sigb;\delta\mathring{\tau},0)$.

Before doing this, a few remarks may be helpful to put things in perspective. The usual massless $\phi^4$ theory in $4-\epsilon$ dimensions  is well known to be IR singular in perturbation theory. IR poles occur at all rational $\epsilon_r=2/r$ with  $r\in\mathbb{N}$ \cite{Sym73,Sym73b,BD82}. Nevertheless, the critical theory exists in the superrenormalizable case $\epsilon>0$. The situation was clarified by Symanzik  who showed that the bare critical mass parameter $\mathring{\tau}_{\mathrm{CP}}$ (the CP analogue of $\mathring{\tau}_{\mathrm{LP}}$) is nonzero, nonanalytic in the coupling constant and not fully accessible by perturbation theory \cite{Sym73,Sym73b}. On dimensional grounds, it can be written as $\mathring{\tau}_{\mathrm{CP}}=
\mathring{u}^{2/\epsilon}\,\mathcal{T}(\epsilon)$, where  $\mathcal{T}(\epsilon)$ is a meromorphic function of $\epsilon=4-d$, with simple poles at $\epsilon_r$. Likewise, one should have
\begin{equation}
\label{eq:tauLPform}
\mathring{\tau}_{\mathrm{LP}}=\mathring{u}^{2/\epsilon}\,\sigb^{-m/2\epsilon}\,\mathcal{T}_m(\epsilon),
\end{equation}
where the functions $\mathcal{T}_m(\epsilon)$ are expected to display analogous nonanalytic behavior in $\epsilon\equiv d^*(m)-d$ at $\epsilon_r$.

In RG studies of CP properties based on the $\epsilon$~expansion the problem of these IR poles is by-passed by choosing $\epsilon$  smaller than the smallest value of the IR poles $\epsilon_r$ occurring at a given order of the calculation. Working at fixed dimensions $d$ requires the use of massive RG scheme (see, e.g.\ \cite{Par80}, \cite{SD89}, \cite[chapter 28.2]{ZJ96} and the extension of this approach to systems with surfaces introduced by two of us \cite{DS94,DS98}). These schemes avoid the perturbative calculation of critical mass parameters such as $\mathring{\tau}_{\mathrm{CP}}$ by expressing the quantities of interest in terms of the bulk correlation length. Crucial prerequisites for their quantitatively successful applications to the study of bulk critical behavior are the availability of perturbation-theory results to sufficiently high orders of the loop expansion \cite{BNM78,AS95} as well as detailed information about the large-order behavior of the required series (for details and references, see e.g.\ \cite{ZJ96} and \cite{PV02}). We are here in a much less favorable situation. First, we are not aware of any reliable quantitative investigations of bulk critical behavior at LP based on the massive field-theory approach, let alone of finite-size effects. (There is, however, some recent work using the so-called exact RG equations \cite{Bervillier2004110}.)  Second, even the application of massive fixed-dimension RG schemes to the study of finite-size effects at CP has just begun \cite{Doh09}. Third, the technical challenges one is faced with in such investigations  of finite-size effects are enormous, even more so for systems with LP. 

Let us therefore take a pragmatic point of view and explore what the theory based on the  smallness parameter $\epsilon$ yields. To this end, we include the shift~\eqref{eq:deltatauren} in the free $\varphi$-propagator of the Hamiltonian~\eqref{eq:Hvarphipart}. 
The contributions to $f_\varphi$ we need to the order of our calculation then are simply given by the free-energy graphs to two-loop order of the ($d-1$)-dimensional action
\begin{equation}
 \label{eq:Fgexpfvarphi}
A\,f_{\varphi}(L)=\,-\,\raisebox{-7pt}{\includegraphics[width=20pt,clip]{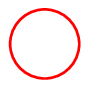}}\,\,-\,\,\raisebox{-12pt}{\includegraphics[width=16pt,clip]{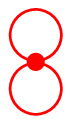}}\,\,-\ldots,
\end{equation}
where the red dot represents the coupling constant $(-\mathring{u}/L)$. Upon exploiting again relation~\eqref{eq:Cmrel}, we can infer the resulting contribution to the residual free energy density $f_{\varphi,\mathrm{res}}(L)$ from equations (4.28)  of \cite{GD08}.  We obtain
\begin{eqnarray}
 \label{eq:fvarphitwoloop}
f_{\varphi,\mathrm{res}}(L)& &=  -n\,\sigb^{-m/4}\,C_m\Bigg[\frac{A_{d-m/2-1}}{d-m/2-1}\, \big(\delta\mathring{\tau}_{L,\|}^{\mathrm{PBC}}\big)^{(2d-m-2)/4}
\nonumber\\ &&\strut 
-\frac{\mu^\epsilon u}{ L}\,\frac{n+2}{4!}\, \frac{A_{d-m/2-1}^{2}}{F_{0,\epsilon}}\, \big(\delta\mathring{\tau}_{L,\|}^{\mathrm{PBC}}\big)^{d-3-m/2}\Bigg]+\ldots,
\end{eqnarray}
where $A_d$ denotes the coefficient defined in equation~\eqref{eq:Ad}, and equation~\eqref{eq:Cmorig}  was used.

The results given in equations~\eqref{eq:deltatauren}--\eqref{eq:fvarphitwoloop} and \eqref{eq:fpsiBC}--\eqref{eq:DeltaPBCCP}  can now be combined in a straightforward fashion and substituted into equation~\eqref{eq:fLfpsifvarphi}.  Evaluating the resulting expression at the fixed-point value~~\eqref{eq:ustar} of the coupling constant $u$ then gives the small-$\epsilon$ expansions~\eqref{eq:DeltaparDD} and \eqref{eq:DeltaparPBC} of $\Delta_\|^{(\mathrm{O,O})}$ and $\Delta_\|^{\mathrm{PBC}}$, respectively. As can be seen from equation~\eqref{eq:fvarphitwoloop},  beyond the order of $\epsilon^{3/2}$, higher half-integer powers appear in conjunction with powers of $\ln\epsilon$.

In \tref{tab:Delpar} we present numerical values for the Casimir amplitudes $\Delta_{\|}^{\mathrm{BC}}(d,m,n)$ of the Gaussian and $n$-vector models with  $d=3$  and $m=0,1,2$. 
\begin{table}[htb]
\centering
\caption{Casimir amplitudes at $d=3$ for parallel slab orientation. $\Delta_{\|,\mathrm{G}}^{\mathrm{BC}}$ denote the Casimir amplitudes of the Gaussian models under  boundary conditions $\mathrm{BC}$. Results obtained for $\phi^{4}$ models by extrapolation of the small-$\epsilon$ expansions to $d=3$ are labelled by subscripts `$\mathrm{extr}$'.}\label{tab:Delpar}
\resizebox{12.5cm}{!}{{
 \begin{tabular}{l . . .}\br 
  \hspace{3em} $m$ & 0 & 1 & 2\\\mr
$\Delta_{\|,\mathrm{G}}^{\mathrm{PBC}}(3,m,n)/n$ & -0.19131 & -0.15792 & -0.13090 \\[0.5em]
   $\Delta_{\|,\mathrm{extr}}^{\mathrm{PBC}}(3,m,1)^{\rm a}$ & -0.19670 & -0.14627 &  -0.088776 \\[0.5em]
   $\Delta_{\|,\mathrm{G}}^{(\mathrm{O,O})}(3,m,n)/n$ & -0.023914 &  -0.027917 &  -0.032725
    \\[0.5em]
$\Delta_{\|,\mathrm{extr}}^{(\mathrm{O,O})}(3,m,1)^{\rm a}$ & -0.011659  &  -0.0076387  & -0.0041162 \\[0.5em]
 $\Delta_{\|,\infty}^{\mathrm{PBC}}(3,m)$ & -0.15305 & -0.12698 &-\\[0.5em]
  $\Delta_{\|,\infty}^{(\mathrm{O,O})}(3,m)$ &  -0.012(1)^{\rm b} & - & -\\
  \br
\end{tabular}}}
\begin{description}\scriptsize
\item[] $^{\rm a}$ Values obtained by evaluating the small-$\epsilon$ expansion at $\epsilon=1+m/2$ ($d=3$).\item[] $^{\rm b}$ Value taken from [38].
\end{description}
\end{table}
We do not consider $(d\,{=}\,3)$-dimensional $\phi^4$ models  with $m=3$ because their lower critical dimension $d_*(m)=2+m/2$ exceeds $3$. However, the Gaussian critical case with $m=d=3$, corresponding to a Hamiltonian with $\mathcal{L}_{\mathrm{b}}=(\Delta\bm{\phi})^2/2$, is meaningful. Therefore, we have included it.

\subsection{Perpendicular slab orientation}\label{sec:CAperp}
In our calculations of Casimir amplitudes for parallel slab orientations we could benefit from a simplifying feature: the required Feynman integrals could be related to their CP analogues. For perpendicular slab orientation, there is no such luxury.  The necessary calculations become more involved, and especially in the case of $(\mathrm{O,O})$ BC, require new techniques. 

We start by decomposing the total reduced free-energy area densities for $\mathrm{PBC}$ and $(\mathrm{O,O})$ BC by analogy with equation~\eqref{eq:fLfpsifvarphi} as 
\begin{eqnarray}
 \label{eq:fperppsipbc}
 f^{\mathrm{BC}}_{\perp}(L)=  f^{\mathrm{BC}}_{\psi;\perp}(L)+\delta_{\mathrm{BC},\mathrm{PBC}}\,f^{\mathrm{BC}}_{\varphi;\perp}(L)
\end{eqnarray}
into their nonzero-mode parts $f^{\mathrm{BC}}_{\psi;\perp}(L)$ and a remaining zero-mode part $f^{\mathrm{PBC}}_{\varphi;\perp}(L)$ that is present only for $\mathrm{BC}=\mathrm{PBC}$. Using a subscript $[l]$ to specify the contribution to $ f^{\mathrm{BC}}_{\psi;\perp}(L)$  of $l$th order in the loop expansion, we write\begin{equation}
 \label{eq:fperppsile}
 f^{\mathrm{BC}}_{\psi;\perp}(L) =f^{\mathrm{BC}}_{\psi;\perp;[1]}(L)+f^{\mathrm{BC}}_{\psi;\perp;[2]}(L)+\Or(3\mbox{-loops}),
 \end{equation}
where the two terms on the right-hand side correspond to the analogues of the graphs displayed in equation~\eqref{eq:fpsigraphs}. 

We consider both BC, $\mathrm{PBC}$ and $(\mathrm{O,O})$, separately, beginning with the technically simpler case of PBC.

\subsubsection{Periodic boundary conditions}\label{sec:CAperpPRB}\strut\\

\begin{figure}[htp]
\centering
\includegraphics[width=30em]{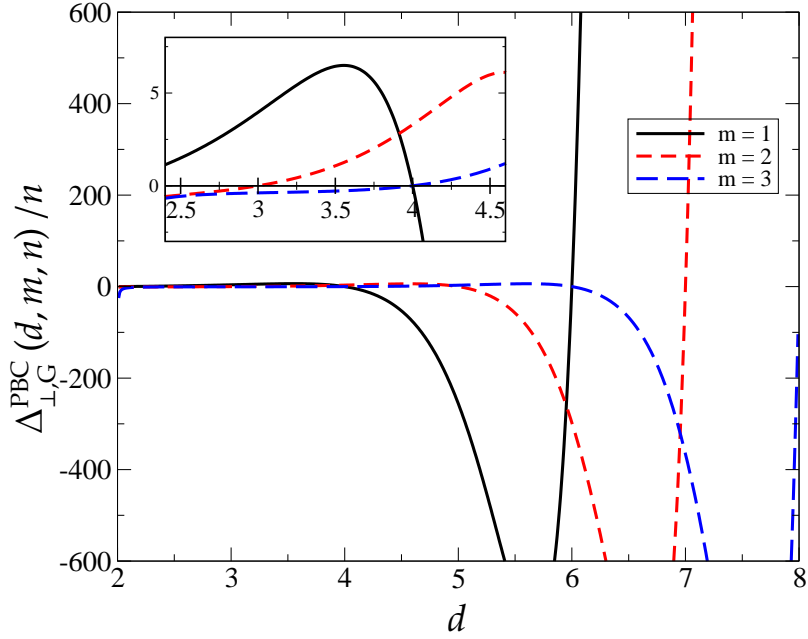}
\caption{Gaussian Casimir amplitudes $\Delta_{\perp,\mathrm{G}}^{\mathrm{PBC}}(d,m,n)/n$ with $m=1,2,3$, plotted as functions of $d$.} \label{fig:Delperp}
\end{figure}

As before, we first consider the Gaussian case $\mathring{u}=0$. Details of the calculation of the one-loop term $f_{\perp,[1]}^{\mathrm{PBC}}(L)$ are presented in  \ref{app:fgperppbc}. The result yields the residual free energy
\begin{equation}
\label{eq:fresperpGauss}
f^{\mathrm{PBC}}_{\mathrm{res},\perp;[1]}(L)\big|_{\tb=\rhob=0}=\sigb^{(d-m)/2}\,\Delta_{\perp,\mathrm{G}}^{\mathrm{PBC}}(d,m,n)\,L^{-(2d-m-1)}
\end{equation}
with the Gaussian Casimir amplitude
\begin{eqnarray}
\label{eq:DelperpPRBGauss}
\frac{\Delta_{\perp,\mathrm{G}}^{\mathrm{PBC}}(d,m,n)}{n}&=&-2\pi P^{(0)}_{m,d+2}\nonumber\\ &=&-\frac{ 2^{2
  d-2 m+1}  \Gamma(d-m/2)\,\zeta (2 d-m)}{\pi ^{(d-1)/2}\,\Gamma[(m-d+1)/2]},
\end{eqnarray}
where we introduced the quantity
\begin{eqnarray}
  \label{eq:Pmd0def}
 P_{m,d}^{(0)}=\frac{ 2^{2
  d-2 m-4}  \Gamma(d-2-m/2)\,\zeta (2 d-m-4)}{\pi ^{(d-1)/2}\,\Gamma[(m-d+3)/2]}
\end{eqnarray}
for subsequent use. In \fref{fig:Delperp} the amplitudes~\eqref{eq:DelperpPRBGauss} with $m=1,2,3$ are plotted as functions of $d$.

Turning to the small-$\epsilon$ expansion, we note that the zero-mode contribution to $f^{\mathrm{PBC}}_{\mathrm{res},\perp}(L)|_{\tb=\rhob=0}$ vanishes at zero-loop order. Hence the one-loop term $f^{\mathrm{PBC}}_{\psi,\mathrm{res},\perp;[1]}(L)|_{\tb=\rhob=0}$ is also given by equation~\eqref{eq:DelperpPRBGauss}. Both the two-loop contribution to $f^{\mathrm{PBC}}_{\psi,\perp}(L)$  and the shift originating  from the graph $\raisebox{-0.5em}{\includegraphics[width=30pt]{./twopteff.eps}}$ can  be computed in a straightforward fashion along lines similar to those followed in \ref{app:fgperppbc}. The results are
\begin{eqnarray}
\label{eq:fpsiperpPRB}\fl
f^{\mathrm{PBC}}_{\psi,\perp;[2]}(L)\Big|_{\tb=\rhob=0}=-\left[\raisebox{-12pt}{\includegraphics[width=16pt,clip]{./twoloopfpsi.eps}}\right]_{\perp}^{\mathrm{PBC}}\nonumber\\
=\mathring{u}\,\sigb^{-m/4}\,\frac{n(n+2)}{4!}\,\sigb^{d-2-3m/4}\,L^{2m-4d+9}\,\left[P^{(0)}_{m,d}\right]^2.
\end{eqnarray}
and
\begin{equation}
\label{eq:deltauperp}
\delta\mathring{\tau}_{L,\perp}^{\mathrm{PBC}}=\mathring{u}\,\sigb^{-m/4}\frac{n(n+2)}{6}\,\left(\sigb^{-1/4}L\right)^{m-2d+4}P_{m,d}^{(0)}.
\end{equation}

\begin{figure}[htbp]
\begin{center}
\includegraphics[width=30em]{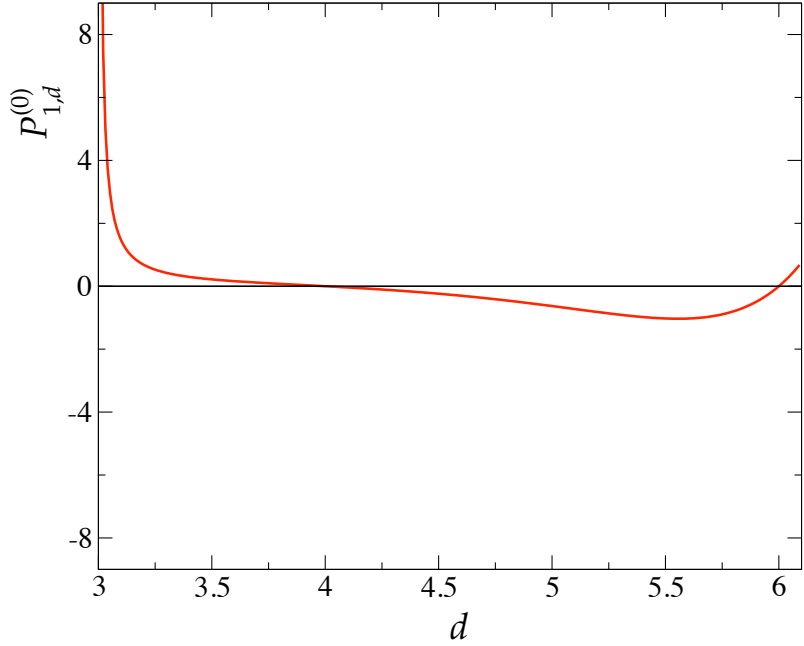}
\caption{$P^{(0)}_{m,d}$, introduced in equation~\eqref{eq:Pmd0def} and used throughout equations~\eqref{eq:fpsiperpPRB}-\eqref{eq:deltauperp}, as a function of $d$ for $m=1$.}
\label{fig:Pmd}
\end{center}
\end{figure}

The quantity $P_{m,d}^{(0)}$ with $m=1$ is plotted as a function of $d$ in \fref{fig:Pmd}. %
Unlike the shift $\delta\mathrm{\tau}_{L,\|}^{\mathrm{PBC}}$ given in equation~\eqref{eq:deltatauren}, this quantity --- and hence the associated $\delta\mathring{\tau}_{L,\perp}^{\mathrm{PBC}}$ --- is negative near the upper critical dimension $d^*(1)=9/2$. In the biaxial case $m=2$, both $P_{m,d}^{(0)}$ and the shift $\delta\mathring{\tau}_{L,\perp}^{\mathrm{PBC}}$ vanish at the upper critical dimension $d^*(2)=5$. A nonpositive shift is a clear warning that a reorganization of RG-improved perturbation theory along the lines used  in \cite{DGS06,GD08,DG09} to deal with the zero-mode problem, and in section~\ref{sec:calcpar} in the case of parallel slab orientation, may not work in the present perpendicular case. Although one might hope that  an appropriate extrapolation of the shift to $d=3$ dimensions would yield a positive value, there is no guarantee that this is the case, nor that a meaningful small $\epsilon$-expansion will result. Unfortunately, we are not able to clarify these issues in a satisfactory fashion. We therefore take a pragmatic point of view, outline our strategy to derive a
small-$\epsilon$ expansion, and show that it fails when $m>0$ because the shift becomes negative for small $\epsilon>0$. 

The Hamiltonian $\mathcal{H}[L^{-1/2}\bm{\varphi}]$ now differs from the expression given in \eqref{eq:Hvarphipart} in that the sum over $\alpha$ is restricted to $\alpha=1,\ldots,m-1$ and the sum over $\beta$  runs over all $d-m$ values $\beta=m+1,\ldots,d$. We make the replacement $\mathring{\tau}\to\mathring{\tau}+\delta\tb^{\mathrm{PBC}}_{L;\perp}$ in this Hamiltonian and set $\mathring{\tau}=0$. Using again relation~\eqref{eq:Cmrel} and remembering that the $\varphi^4$ coupling constant of this effective $(d-1)$-dimensional bulk action is given by $\mathring{u}/L$, we obtain
\begin{eqnarray}
\label{eq:fvarphiresPRB}\fl
f^{\mathrm{PBC}}_{\varphi,\mathrm{res};\perp}(L)= -n\,\sigb^{-m/4+1/4}\,C_{m-1}\Bigg[\frac{A_{d-m/2-1/2}}{d-m/2-1/2}\big(\delta\mathring{\tau}_{L,\perp}^{\mathrm{PBC}}\big)^{(2d-m-1)/4}\nonumber\\
\strut -\frac{\mu^\epsilon u}{L}\,\frac{(n+2)}{4!}\,\frac{A^2_{d-m/2-1/2}}{F_{0,\epsilon}}\big(\delta\mathring{\tau}_{L,\perp}^{\mathrm{PBC}}\big)^{(2d-m-5)/2}\Bigg]+\ldots,
\end{eqnarray}
where $C_m$ and $A_d$ were defined in equations~\eqref{eq:Cm} and \eqref{eq:Ad}, respectively. To compute the small-$\epsilon$ expansion of the Casimir amplitude $\Delta_{\perp}^{\mathrm{PBC}}$, we must add $f^{\mathrm{PBC}}_{\varphi,\mathrm{res};\perp}(L)$ to the two-loop results for $f^{\mathrm{PBC}}_{\psi,\mathrm{res};\perp}(L)$ derived above, express the result in terms of renormalized quantities and evaluate it at $u=u^*$, $\tau=\rho=0$, and $\sigma=\mu=1$. Since the shift $\delta\mathring{\tau}_{L,\perp}^{\mathrm{PBC}}\sim\epsilon$ (in our attempted approximation), the first term in equation~\eqref{eq:fvarphiresPRB} varies as $\epsilon^{7/4}$. Hence we expect a fractional power series expansion of the form
\begin{equation}
\label{eq:DelperpPBCepsexp}
\frac{\Delta_{\perp}^{\mathrm{PBC}}(d,m,n)}{n}=a_0(m)+a_1(m,n)\epsilon+a_{7/4}(m,n)\epsilon^{7/4}+\mathrm{o}(\epsilon^{7/4}),
\end{equation}
whose higher-order terms should again be modified by powers of $\ln \epsilon$.

The terms of zeroth and first order in $\epsilon$ of this series originate entirely from $f^{\mathrm{PBC}}_{\psi,\mathrm{res};\perp}(L)$. The associated series coefficients $a_0$ and $a_1$ are well defined and given by
\begin{equation}
\label{eq:a0}
a_0(m)=-\frac{2^{9-m}\pi^{(26-m)/4}}{1575\,\Gamma[(m-6)/4]}
\end{equation}
and
\begin{eqnarray}
 \label{eq:a1}\fl
a_1(m,n)= \frac{2^{9-m}\pi^{(26-m)/4}\left\{11-6\gamma_E+\ln(2^{12}\pi^{-3})+3\,\psi[(m-6)/4]+12\,\zeta'(8)/\zeta(8)\right\}}{9450\,\Gamma(m/4-3/2)}\nonumber\\  \strut +\frac{2^{5-m}\pi^{(26-m)/4}\,(m-2)}{2025\,\Gamma(m/4-1/2)}\,\frac{n+2}{n+8}\,.
\end{eqnarray}
For the coefficient $a_{7/4}(m,n)$, one finds
\begin{equation}
 \label{eq:a7ov4}\fl
a_{7/4}(m,n)= -\frac{2^{-m+7}\pi^{6-m/4}\Gamma(-7/4)\,\left[(m-2)/4\right]^{7/4}}{3^{7/2}5^{7/4}\,\Gamma[(m+1)/4]}\,\left(\frac{n+2}{n+8}\right)^{7/4}.
\end{equation}
In the uniaxial case $m=1$, it becomes complex, which renders the expansion useless. This is a direct consequence of the fact that $\delta\mathring{\tau}_{L,\perp}^{\mathrm{PBC}}$ approaches a negative number as $\epsilon\to 0+$ when $m=1$. 

\subsubsection{Large-\texorpdfstring{$n$}{n} results for perpendicular orientation and periodic boundary conditions}

As an alternative to the small-$\epsilon$ expansion, we now wish to use the large-$n$ limit to gain information about the Casimir amplitude $\Delta_{\perp}^{\mathrm{PBC}}$ for the uniaxial case in $d=3$ dimensions. The result could be derived by solving a corresponding mean spherical model with short-range interactions along the lines followed in \cite{DDG06}. However, the large-$n$ limit of the $\phi^4$~model we worked with throughout this paper  (cf.\ section~\ref{sec:Mod}) can also be obtained directly. Since we are interested in the solution at the bulk LP, we set $\mathring{\tau}$ and $\mathring{\rho}$ to their bulk LP values $\mathring{\tau}_{\mathrm{LP}}$ given in equation~\eqref{eq:tauLPtauCP} and $\mathring{\rho}_{\mathrm{LP}}=0$ and look for $O(n)$ symmetric solutions. Let us write the self-energy as $-\vartheta^{\mathrm{PBC}}_{\perp,L}$. In the large-$n$ limit, one has $\vartheta^{\mathrm{PBC}}_{\perp,L}=\mathring{u}n\langle \phi_1^2\rangle/6$ where the latter average must be computed with the self-consistent propagator involving the self-energy $-\vartheta^{\mathrm{PBC}}_{\perp,L}$ (see e.g.\ \cite{ZJ96}). To obtain a meaningful  $n\to \infty$ limit, we must scale $\mathring{u}$ by $n$, keeping
\begin{equation}
\label{eq:gdef}
\mathring{g}=n\mathring{u}
\end{equation}
fixed. 

We can now subtract the self-consistent equation for $\vartheta^{\mathrm{PBC}}_{\perp,L}$ and its  analogue for the bulk quantity $\vartheta_{\mathrm{LP}}\equiv\vartheta^{\mathrm{PBC}}_{\mathrm{LP}}=\lim_{L\to\infty} \vartheta^{\mathrm{PBC}}_{\perp,L}$ at the LP to obtain a self-consistent equation for the difference $\delta\vartheta^{\mathrm{PBC}}_{\perp,L}\equiv\vartheta^{\mathrm{PBC}}_{\perp,L}-\vartheta^{\mathrm{PBC}}_{\mathrm{LP}}$. The bulk quantity $\vartheta^{\mathrm{PBC}}_{\mathrm{LP}}$ does not, of course, depend on whether the surface orientation is perpendicular or parallel. Furthermore, we can use the fact that the self-consistent bulk propagator is massless at the LP. We thus arrive at the equation
\begin{eqnarray}\label{eq:sceperplargen}
\fl
\frac{6}{\mathring{g}}\delta\vartheta^{\mathrm{PBC}}_{\perp,L}=\int_{\bm{k}}^{(m-1)}\int_{\bm{p}}^{(d-m)}\Bigg[\frac{1}{L}\sum_{r=-\infty}^\infty \frac{1}{p^2+\sigb\,[k^2+(2\pi r/L)^2]^2+\delta\vartheta^{\mathrm{PBC}}_{\perp,L}}
\nonumber\\ \qquad\strut
 -\int_{-\infty}^\infty\frac{\rmd{K}}{2\pi}\,\frac{1}{p^2+\sigb(k^2+K^2)^2}\Bigg],
\end{eqnarray}
whose analogue for parallel orientation, \eqref{eq:spteqPBC},  is used in \ref{app:largen}.

The significance of the quantity $\delta\vartheta^{\mathrm{PBC}}_{\perp,L}(d,m;\mathring{g})$ should be clear: it
is the finite-size susceptibility, taken at the LP. We make it dimensionless, introducing
\begin{equation}
\mathcal{R}_\perp= \delta\vartheta^{\mathrm{PBC}}_{\perp,L}L^4/\mathring{\sigma},
\end{equation}
and then use Poisson's summation formula
\begin{equation}\label{eq:Poisson}
\frac{1}{L}\sum_{r=-\infty}^\infty f(2\pi r/L)=\sum_{j=-\infty}^\infty\int_{-\infty}^\infty\frac{dK}{2\pi}f(K)\,e^{iKjL}
\end{equation}
to express the right-hand side of equation~\eqref{eq:sceperplargen} in terms of the free propagator $G_{\mathrm{b}}^{(d,m)}$. To this end, we define a function
\begin{eqnarray}\label{eq:Qperpdef}
\fl
Q^\perp_{d,m}(y)&\equiv&\frac{y}{2}\int_{\bm{k}}^{(m-1)}\int_{\bm{p}}^{(d-m)}\left[\sum_{K\in 2\pi\mathbb{Z}}-\int_{-\infty}^\infty\frac{dK}{2\pi}\right]\frac{1}{p^2+(k^2+K^2)^2+y}\nonumber\\
\fl
&=&y\sum_{j=1}^\infty G_{\mathrm{b}}^{(d,m)}(j\vec{e}_z|\mathring{\sigma}=1;y,\mathring{\rho}=0)
\end{eqnarray} 
by analogy with equation~\eqref{eq:Qd2def}, whose representation in the second line follows via the summation formula~\eqref{eq:Poisson}. Evaluated at $L=\mathring{\sigma}=1$, the right-hand side of equation~\eqref{eq:sceperplargen} then becomes 
\begin{equation}\label{eq:rhssce}
\frac{2\,Q^\perp_{d,m}(\mathcal{R}_\perp)}{\mathcal{R}_\perp}-\int_{\bm{k}}^{(m)}\int_{\bm{p}}^{(d-m)}\frac{\mathcal{R}_\perp}{(p^2+k^4+\mathcal{R}_\perp)(p^2+k^4)}.
\end{equation}

The dimensionless susceptibility $\mathcal{R}_\perp$ approaches a nontrivial number independent of $L$ as $L\to\infty$. This number is the zero of the function~\eqref{eq:rhssce}. The term on the left-hand side of the self-consistency equation~\eqref{eq:sceperplargen} gives corrections to scaling, which we ignore. Upon evaluation of the bulk integral in expression~\eqref{eq:rhssce}  we thus arrive at the self-consistency condition 
\begin{equation}\label{eq:sccRperp}
-C_m\,A_{d-m/2}\,\mathcal{R}_\perp^{(d-m/2-2)/2}+2\mathcal{R}_\perp^{-1}\,Q_{d,m}^\perp(\mathcal{R}_\perp)=0
\end{equation}
for $\mathcal{R}_\perp$. This applies to those values of $d$ and $m$ for which the required integrals and the function $Q_{d,m}^\perp$ are well-defined.  This is the case if we require that $2<d-m/2<4$. For $d=3$ dimensions, we are left with the interesting uniaxial case $m=1$. The required bulk propagator
\begin{eqnarray}\label{eq:Gb31}
G_{\mathrm{b}}^{(3,1)}(z\vec{e}_z|1;y,0)&=&\int_{\bm{k}}^{(1)}\int_{\bm{p}}^{(d-1)}\frac{e^{ikz}}{p^2+k^4+y}\nonumber\\ &=&\frac{1}{2\pi|z|}\,\cos{\big(zy^{1/4}/\sqrt{2}\big)}\,e^{-|z|y^{1/4}/\sqrt{2}}
\end{eqnarray}
can be calculated in a straightforward fashion. Inserting it into the last line of equation~\eqref{eq:Qperpdef} and performing the summation gives
\begin{equation}\label{eq:Q31perp}
Q_{3,1}^\perp(y)=-\frac{y}{4\pi}\,\ln\Big[1+e^{-(4y)^{1/4}}-2\,e^{-(y/4)^{1/4}}\cos(y/4)^{1/4}\Big].
\end{equation}
It is now an easy matter to determine the root $\mathcal{R}_\perp(d,m)$ of equation~\eqref{eq:sccRperp} for $(d,m)=(3,1)$ by numerical means using {\sc Mathematica} \cite{Mathematica7}. One obtains
\begin{equation}
\mathcal{R}_\perp(3,1)=0.998614\ldots.
\end{equation}
Thus the large-$n$ limit yields a well-defined positive value for the scaled inverse finite-size susceptibility $\mathcal{R}_\perp$ at $(d,m)=(3,1)$, unlike the $\epsilon$ expansion.

To determine the Casimir amplitude $\Delta^{\mathrm{PBC}}_{\perp,\infty}$ we must compute
\begin{equation}
\Delta^{\mathrm{PBC}}_{\perp,\infty}=\lim_{L\to\infty}L^{2d-m-1}f_{\mathrm{res},\mathrm{LP},\infty}|_{\mathring{\sigma}=1},
\end{equation}
where the subscript $\infty$ again indicates that the limit $f_{\mathrm{res},\mathrm{LP},\infty}=\lim_{n\to\infty} f_{\mathrm{res},\mathrm{LP}}/n$ has been taken. This limit is given by the value  $D(\mathcal{R}_\perp)$ of a function $D(y)$ that satisfies $D(0)=
\Delta_{\perp,\mathrm{G}}^{\mathrm{PBC}}(d,m,1)$ and can be determined by integrating the equation of state~\eqref{eq:sccRperp} with respect to $\mathcal{R}_\perp$. Hence we have
\begin{equation}\fl
\Delta^{\mathrm{PBC}}_{\perp,\infty}(d,m)=-\frac{A_{d-m/2}}{d-m/2}\,C_m\,\mathcal{R}_\perp^{(2d-m)/4}+\Delta_{\perp,\mathrm{G}}^{\mathrm{PBC}}(d,m,1)+\int_0^{\mathcal{R}_\perp}\frac{dy}{y}\,Q^\perp_{d,m}(y).
\end{equation}
We now set $(d,m)=(3,1)$, insert $Q^\perp_{3,1}(y)$ from equation~\eqref{eq:Q31perp}, and use {\sc Ma\-the\-ma\-ti\-ca} \cite{Mathematica7} to perform the integral. This gives, after some rearrangements,
\begin{eqnarray}\label{eq:Delperpprbinfty31}\fl
\Delta^{\mathrm{PBC}}_{\perp,\infty}(3,1)=-\frac{y^{5/4}}{5\pi \sqrt{2}  }-\frac{\sqrt{2} }{\pi}\mathrm{Re}\bigg\{(1-i) y^{3/4}\, \mathrm{Li}_2{\left[e^{-(1+i)
  (y/4)^{1/4}}\right]}\nonumber\\  \strut
  -3 i \sqrt{2} y^{1/2}\,
   \mathrm{Li}_3{\left[e^{-(1+i) (y/4)^{1/4}}\right]} -(6+6 i)
   y^{1/4}\, \mathrm{Li}_4{\left[e^{-(1+i)
  (y/4)^{1/4}}\right]}\nonumber\\ \strut
  -6 \sqrt{2} \,\mathrm{Li}_5{\left[e^{-(1+i)
  (y/4)^{1/4}}\right]}\bigg\}\bigg|_{y=\mathcal{R}_\perp(3,1)}
\nonumber\\  \simeq 4.00053,
\end{eqnarray}
where $\mathrm{Li}_s(z)$ is the polylogarithm, defined by \cite{NIST-DLMF2010_25.12}
\begin{equation}
\mathrm{Li}_s(z)=\sum_{j=1}^\infty\frac{z^j}{j^s}.
\end{equation}

Note that the sign of this Casimir amplitude is positive. That is, the Casimir force is \emph{repulsive} in this case. This is remarkable because, under PBC, the corresponding Casimir forces  at uniaxial LPs and parallel orientation  and at  CP are both attractive [cf.\ equation~\eqref{eq:largenpar} and table~\ref {tab:Delpar}, on the one hand, and equation~\eqref{eq:CPlargenpar}, on the other hand]. More generally, there exist theorems \cite{KK06,Bac07} which state that  Casimir forces are guaranteed to be negative (attractive) when certain symmetry properties (such as the same BC on both planes and reflection-positive interactions) are satisfied. In the LP case with perpendicular orientation, reflection positivity is violated along the $z$-axis due to the presence of competing nearest and axial next-nearest neighbour interactions along this direction. Thus the theorem of \cite{Bac07} does not apply. The one of \cite{KK06} is restricted to Gaussian models. It also  does not apply to our LP case, neither on  the level of the corresponding Gaussian model nor in the large-$n$ limit, because it presumes a Gaussian action involving square-gradient terms of the order parameter but no higher-order derivatives.

\subsubsection{Ordinary-ordinary boundary conditions}\label{sec:CAperpord}\strut\\

The case of perpendicular orientation with $(\mathrm{O,O})$ BC turns out to be technically very demanding. For the sake of simplicity, we will therefore content ourselves here with an analysis to one-loop order.

To compute $f_{\mathrm{res},\perp;[1]}^{(\mathrm{O,O})}(L)$, we proceed as follows. We consider the Gaussian infinite-space model with Hamiltonian
\begin{equation}
\mathcal{H}_{\mathrm{G}}=\int_{\mathbb{R}^d}\rmd^dx\,\mathcal{L}_{\mathrm{b}}(\bm{x})\big|_{\mathring{u}=0;\tb=\rhob=0}.
\end{equation}
 We now insert two defect planes into the system, one at $z=0$ and a second at $z=L$, where $z$ is taken to be an $\alpha$-direction. At either one of them the BC~\eqref{eq:bcasperp} are required to hold. We then compute the change in reduced free energy $\Delta F_{\mathrm{G}}(k_BT)^{-1}$ caused by the defect planes. The $L$-dependent part of the area density $\lim_{A\to\infty}F_{\mathrm{G}}(k_BTA)^{-1}$ gives us the desired  residual reduced free-energy density $f_{\mathrm{res},\perp;[1]}^{(\mathrm{O,O})}(L)$. Note that this conclusion hinges on the BC~\eqref{eq:bcasperp}, which ensures that the region between the defect planes is decoupled from the outside regions $z<0$ and $z>L$.\footnote{In general, systems with two defect planes must be carefully distinguished from slabs confined by two parallel plates at a distance $L$. A decoupling analogous to the one just mentioned  occurs in the case of the standard infinite-space $\phi^4$ model in the presence of two defect planes at which Dirichlet BC hold \cite{Sym81}.  Dirichlet BC would also suffice to achieve this decoupling in the LP case if the defect planes were oriented perpendicular to a $\beta$-direction. For a detailed discussion of the similarities and differences of boundary critical behavior at free surfaces planes and defect planes, the reader might want to consult  \cite{BE81}, \cite{DDE83} and \cite[chapter IV.A.4]{Die86a}.}
 
We implement the BC~\eqref{eq:bcasperp} at the planes $\mathfrak{B}_1$ ($z=0$) and $\mathfrak{B}_2$ ($z=L$) in the required functional integral by delta functions. Using a standard trick \cite{FaddeevPopov67,LK92}, we represent these constraints through  functional integrals over four $n$-component fields $\bm{\vartheta}_j$ whose supports are restricted to $\mathfrak{B}_1$ and  $\mathfrak{B}_2$ for $j=1,2$ and $j=3,4$, respectively.  Thus 
\begin{equation}\fl
\prod_{\bm{x}_1\in\mathfrak{B}_1}\delta[\bm{\phi}(\bm{x}_1)]\,\delta[\partial_{z_1}\bm{\phi}(\bm{x}_1)]\prod_{\bm{x}_2\in\mathfrak{B}_2}\delta[\bm{\phi}(\bm{x}_2)]\,\delta[\partial_{z_2}\bm{\phi}(\bm{x}_2)]
\equiv \int\mathcal{D}[\bm{\vartheta}]\,\rme^{-\rmi \mathcal{C}[\bm{\phi},\bm{\vartheta}]}
\end{equation}
with
\begin{equation}
\mathcal{C}[\bm{\phi},\bm{\vartheta}]=\int_{\mathfrak{B}_1}\left[\bm{\vartheta}_1\cdot\bm{\phi}+\bm{\vartheta}_2\cdot\partial_{z}\bm{\phi}\right]+\int_{\mathfrak{B}_2}\left[\bm{\vartheta}_3\cdot\bm{\phi}+\bm{\vartheta}_4\cdot\partial_z\bm{\phi}
\right],
\end{equation}
where  the functional integration $\int\mathcal{D}[\bm{\vartheta}]$ is  over all four fields $\bm{\vartheta}_j$.

As a straightforward consequence we obtain
\begin{eqnarray}
\rme^{-\Delta F_{\mathrm{G}}/k_BT}&=&\int\mathcal{D}[\bm{\vartheta}]\frac{\int\mathcal{D}[\bm{\phi}]\,\exp\left(-\mathcal{H}[\bm{\phi}]-\rmi\mathcal{C}[\bm{\phi},\bm{\vartheta}]\right)}{\int\mathcal{D}[\bm{\phi}]\,\exp\left(-\mathcal{H}[\bm{\phi}]\right)}\nonumber\\ &=&\int\mathcal{D}[\bm{\vartheta}]\exp\left[-\frac{1}{2}\left\langle(\mathcal{C}[\bm{\phi},\bm{\vartheta}])^2\right\rangle_{\mathcal{H}_{\mathrm{G}}}\right],
\end{eqnarray}
where $\langle\ldots\rangle_{\mathcal{H}_{\mathrm{G}}}$ denotes an average over $\bm{\phi}$ with the path probabililty density $\propto \rme^{-\mathcal{H}_{\mathrm{G}}}\mathcal{D}[\bm{\phi}]$. The remaining Gaussian functional integration over $\bm{\vartheta}$ is straightforward, giving
\begin{eqnarray}
f_{\mathrm{G}}(L)=\lim_{A\to\infty}\frac{\Delta F_{\mathrm{G}}}{Ak_BT}=\frac{n}{2}\left(\mathrm{const}+\lim_{A\to\infty}A^{-1}\ln\det\bm{\mathcal{M}}_L\right),
\end{eqnarray}
where ``const'' is a constant which will drop out in the residual free energy
\begin{equation}
\label{eq:fGresdef}
f_{\mathrm{G,res}}(L)=f_{\mathrm{G}}(L)-f_{\mathrm{G}}(\infty)
\end{equation}
we are interested in. The matrix $\bm{\mathcal{M}}_L$ can be written in a suggestive manner as
\begin{eqnarray}\label{eq:Mmatrix}\fl
\bm{\mathcal{M}}_L=\left(
\begin{array}{cccc}
{_1}|G_{\mathrm{b}}^{(d,m)}|_1 & {_1}|G_{\mathrm{b}}^{(d,m)}\overleftarrow{\partial}_z|_1& {_1}|G_{\mathrm{b}}^{(d,m)}|_2&{_1}|G_{\mathrm{b}}^{(d,m)}\overleftarrow{\partial}_z|_2\\
{_1}|\overrightarrow{\partial}_zG_{\mathrm{b}}^{(d,m)}|_1 & {_1}|\overrightarrow{\partial}_zG_{\mathrm{b}}^{(d,m)}\overleftarrow{\partial}_z|_1& {_1}|\overrightarrow{\partial}_zG_{\mathrm{b}}^{(d,m)}|_2&{_1}|\overrightarrow{\partial}_zG_{\mathrm{b}}^{(d,m)}\overleftarrow{\partial}_z|_2\\
{_2}|G_{\mathrm{b}}^{(d,m)}|_1 & {_2}|G_{\mathrm{b}}^{(d,m)}\overleftarrow{\partial}_z|_1& {_2}|G_{\mathrm{b}}^{(d,m)}|_2&{_2}|G_{\mathrm{b}}^{(d,m)}\overleftarrow{\partial}_z|_2\\
{_2}|\overrightarrow{\partial}_zG_{\mathrm{b}}^{(d,m)}|_1 & {_2}|\overrightarrow{\partial}_zG_{\mathrm{b}}^{(d,m)}\overleftarrow{\partial}_z|_1& {_2}|\overrightarrow{\partial}_zG_{\mathrm{b}}^{(d,m)}|_2&{_2}|\overrightarrow{\partial}_zG_{\mathrm{b}}^{(d,m)}\overleftarrow{\partial}_z|_2
\end{array}
\right).\nonumber\\
\end{eqnarray}
It consists of blocks involving the free bulk propagator $G_{\mathrm{b}}^{(d,m)}$ and its derivatives at points on the defect planes $\mathcal{B}_j$. The derivatives $\overleftarrow{\partial}_z$ and $\overrightarrow{\partial}_z$ act to the left and right, respectively. Further, the notations ${_j}|$ and $|_{j'}$ indicate that the left point  $\bm{x}_1=(\bm{y}_2,z_1)$ lies on $\mathfrak{B}_j $ and the right point $\bm{x}_2=(\bm{y}_2,z_2)$ is located on $\mathfrak{B}_{j'}$. For example, the elements of the block $ {_1}|\overrightarrow{\partial}_zG_{\mathrm{b}}^{(d,m)}\overleftarrow{\partial}|_2$ (labelled by $\bm{y}_1$ and $\bm{y}_2$) are  $\partial_{z_1}\partial_{z_{2}}G_{\mathrm{b}}^{(d,m)}[(\bm{y}_{1},z_1{=}0),(\bm{y}_{2},z_2{=}L)]$.

To exploit  translation invariance along the $\bm{y}$-direction, we Fourier transform with respect to $\bm{y}$. Let $(\bm{k},\bm{p})$ with $\bm{k}\in\mathbb{R}^{m-1}$ and $\bm{p}\in\mathbb{R}^{d-m}$ be the wave-vector conjugate to $\bm{y}$. In this momentum representation, the matrix  $\bm{\mathcal{M}}_L$ is block diagonal. The logarithm of its determinant becomes
\begin{equation}\label{eq:lndetMtilde}
\lim_{A\to\infty}A^{-1}\ln\det \bm{\mathcal{M}}_L=\int_{\bm{k}}^{(m-1)}\int_{\bm{p}}^{(d-m)}\ln\det\widetilde{\bm{\mathcal{M}}}_L(\bm{k},\bm{p}),
\end{equation}
where $\widetilde{\bm{\mathcal{M}}}_L(\bm{k},\bm{p})$ is a $4\times 4$ matrix. To compute its elements, it is helpful to know
\begin{equation}
G^{(d,m)}_{\mathrm{b}}(\bm{k},\bm{p};z_1-z_2)=\int_{-\infty}^\infty\frac{\rmd{K}}{2\pi}\,\frac{\exp\left(\rmi K(z_1-z_2)\right)}{p^2+\sigb(k^2+K^2)^2}\,,
\end{equation}
the free bulk propagator in the $\bm{k}\bm{p}z$-representation. It may be gleaned from  equations (20)--(23) of \cite{DSP06}; we have
\begin{equation}
\label{eq:Gbkpz}\fl
G^{(d,m)}_{\mathrm{b}}(\bm{k},\bm{p};z)=\frac{\kappa_-\cos(\kappa_-\sigb^{-1/4}|z|)+\kappa_+\sin(\kappa_-\sigb^{-1/4}|z|)}
{4\sigb^{1/4}\kappa_-\kappa_+(\kappa_-^2+\kappa_+^2)}\,\rme^{-\kappa_+\sigb^{-1/4}|z|}
\end{equation}
with
\begin{equation}
 \label{eq:kappamp}
\kappa_\mp\equiv\frac{1}{\sqrt 2}\sqrt{\sqrt{p^2+\sigb\,k^4}\mp \sigb^{1/2}\,k^2}\,.
\end{equation}
A straightforward calculation then yields
\begin{equation}
\label{eq:detMkp}\fl
\det \widetilde{\bm{\mathcal{M}}}_L(\bm{k},\bm{p})=
\frac{\kappa_-^2 \cosh (2L\sigb^{-1/4}\kappa_+) + \kappa_+^2 \mbox{cos} (2L\sigb^{-1/4}\kappa_-)
-\kappa_-^2-\kappa_+^2}{128\,\sigb^2\kappa_-^2\kappa_+^4(\kappa_-^2+\kappa_+^2)^2\exp(2L\sigb^{-1/4}\kappa_+)}\,.
\end{equation}

Noting that 
\begin{equation}
\label{eq:lnMinfty}
\lim_{L\to\infty}\det\widetilde{\bm{\mathcal{M}}}_L(\bm{k},\bm{p})=\left[256\,\sigb^2\kappa_+^4(\kappa_-^2+\kappa_+^2)^2\right]^{-1},
\end{equation}
we can make the subtraction in equation~\eqref{eq:fGresdef} to obtain 
\begin{eqnarray}
\label{eq:fGres}\fl
f_{\mathrm{G,res}}(L)\nonumber\\ \fl=\frac{n}{2}\int_{\bm{p}}^{(d-m)}\int_{\bm{k}}^{(m-1)}\ln \bigg[\frac{\kappa_-^2\cosh(2\sigb^{-1/4}L\kappa_+)+\kappa_+^2\cos(2\sigb^{-1/4}L\kappa_-)-\kappa_-^2-\kappa_+^2}{(\kappa_-^2/2)\exp(2\sigb^{-1/4}L\kappa_+)}\bigg].\nonumber\\
\end{eqnarray}
Assuming that $m>1$ and $d-m>0$, we perform the angular integrations. This leaves us with the two radial integrals over $p$ and $k$. To simplify them, we make the change of variables
\begin{equation}
k=\frac{\varkappa q}{\sqrt{2}\,w_\varkappa L}\,,\quad p=\frac{q^2\sigb^{1/2}}{2w_\varkappa^2L^2}\,,\quad\rmd{k}\,\rmd{p}\,=\frac{q^2\sqrt{\sigb/2}}{w_\varkappa^3L^3}\,\rmd{\varkappa}\,\rmd{q},
\end{equation}
where
\begin{equation}
\label{eq:WKdef}
w_\varkappa=\sqrt{\varkappa^2+\sqrt{\varkappa^4+1}}\,.
\end{equation}
We obtain
\begin{equation}
\label{eq:fGresDel}
f_{\mathrm{G,res}}(L)=\sigma^{(d-m)/2}\,\Delta^{(\mathrm{O,O})}_{\mathrm{G},\perp}(d,m,n)\,L^{1+m-2d}
\end{equation}
with the Casimir amplitudes
\begin{eqnarray}
\label{eq:DeltaGperp}\fl
\Delta^{(\mathrm{O,O})}_{\mathrm{G},\perp}(d,m,n)/n=&\frac{K_{d-m}K_{m-1}}{2^{(2d-m-1)/2}}\int_0^\infty\rmd{\varkappa}\,\frac{\varkappa^{m-2}}{w_\varkappa^{2d-m-1}}\int_0^\infty\rmd{q}\,q^{2d-m-2}\nonumber\\  &\times \ln
\left(2\rme^{-q}\left\{
\cosh(q)-1+w_\varkappa^4\big[\cos(q/w_\varkappa^2)-1\big]\right\}\right).
\end{eqnarray}
Since $\sigb=\sigma$ when $\mathring{u}=0$, we have replaced $\sigb$ by $\sigma$ in equation~\eqref{eq:fGresDel}.

This is our final result for $\Delta^{(\mathrm{O,O})}_{\mathrm{G},\perp}$ when $m>1$ and $d>m$. In the special cases $m=1$ (uniaxial LP) and $d=m$ (isotropic LP), the result simplifies to single integrals; one finds
\begin{equation}\label{eq:meq1}\fl
\Delta^{(\mathrm{O,O})}_{\mathrm{G},\perp}(d,1,n)/n=K_{d-1}\,2^{1-d}\int_0^\infty \rmd{q}\,q^{2d-3}
\ln\left\{2\,\rme^{-q}\left[\cosh (q)+ \cos (q)-2\right]\right\}
\end{equation}
and
\begin{equation}\fl\label{eq:meqd}
\Delta^{(\mathrm{O,O})}_{\mathrm{G},\perp}(d,d,n)/n=2^{-d}\,K_{d-1}\int_0^\infty \rmd{q}\,q^{d-2}
\ln\left\{\rme^{-q}\left[2\cosh (q)-2-q^2\right]\right\}.
\end{equation}

The simplest way to get these results is by going back to the appropriate analogues of equation~\eqref{eq:fGres}. To obtain the analogue for the uniaxial case  from this equation, one must drop the integral $\int_{\bm{k}}^{(m-1)}$  and replace the integrand by its $\bm{k}\to\bm{0}$ limit; likewise, one must omit $\int_{\bm{p}}^{(d-m)}$ and take the limit $\bm{p}\to\bm{0}$ of the integrand in the case of the isotropic LP. The result~\eqref{eq:meq1} for the uniaxial case can also be inferred directly from the general one, equation \eqref{eq:DeltaGperp}, with the aid of the formula (see e.g.\ \cite{SD01})
\begin{equation}\label{eq:limDzero}
\lim_{D\to0}K_D\int_0^\infty \rmd{\varkappa}\,\varkappa^{D-1}f_D(\varkappa)=\lim_{D\to0}\int_{\bm{\varkappa}}^{(D)}f_D(\varkappa)=f_0(0),
\end{equation}
where $f_D$ is a rotationally invariant function of $\bm{\varkappa}\in\mathbb{R}^D$ that depends parametrically on $D$. This works because the original limit $k\to0$ at fixed $p$ translates into the limit $\varkappa\to 0$. Deriving the result~\eqref{eq:meqd} for the isotropic LP from equation~\eqref{eq:DeltaGperp} is possible, but more subtle. (One would have to take the appropriate scaling limit $q\sim p$ and $\varkappa\sim p^{-1/2}$ corresponding to the original $\lim_{p\to 0}$ of the integrand.)

In \tref{tab:Deltaperpval} we assembled various values of $\Delta^{(\mathrm{O,O})}_{\mathrm{G},\perp}(d,m,n)/n$ for different $m$ at the corresponding upper critical dimension $d^*=4+m/2$ and for general $m\leq d\leq 8$. For the case of $d=3$, we included all values for $m\leq 3$, in particular the special ones  for the uniaxial $(m=1)$ and the isotropic $(m=3)$ LP.

\begin{table}[htb]
\caption{Casimir amplitudes $\Delta^{(\mathrm{O,O})}_{\mathrm{G},\perp}$ for various values of $d$ and $m$.}\label{tab:Deltaperpval}
\centering
{\Large
\resizebox{10cm}{!}{
\begin{tabular}{lcccc}\br
\hspace{3em} m & 1 & 2 & 3 & 4\\\mr
$\Delta^{(\mathrm{O,O})}_{\mathrm{G},\perp}(d^{*},m,n)/n$ & \phantom{0}-6.596 & \phantom{0}-2.841 & \phantom{0}-1.271 & \phantom{00}-0.5635 \\[0.5em]
 $\Delta^{(\mathrm{O,O})}_{\mathrm{G},\perp}(d^{*},m+4,n)/n$   & \phantom{00}-0.2411 & \phantom{000}-0.09856 & \phantom{000}-0.03848 & \phantom{000}-0.01438 \\[0.5em]
$\Delta^{(\mathrm{O,O})}_{\mathrm{G},\perp}(3,m,n)/n$ & \phantom{00}-1.3240 & \phantom{00}-0.5900 & \phantom{00}-0.2732 & ---  \\%
 \br
\end{tabular}
}}
\end{table}

\section{Summary and conclusions}\label{sec:concl}

In this paper we studied the effects of confining anisotropic scale-invariant systems between two parallel planes at distance $L$. Just as in the case of confined isotropic scale-invariant systems, the confinement of long wave-length fluctuations induces long-ranged effective forces. However, in the anisotropic case, the problem is much richer. Important qualitative and quantitative differences arise. 

A first important qualitative difference implied by the distinct scaling behavior of coordinate separations along $\alpha$- and $\beta$-directions is that the orientation of the boundary planes matters. The exponents $\zeta_{\perp}$ and $\zeta_\|$ characterizing the algebraic decay of the fluctuation-induced interactions between the boundary planes as a function of $L$ differ, depending on whether the surface planes are aligned perpendicular to an $\alpha$-direction or parallel to all $\alpha$-directions, i.e.\ perpendicular to a $\beta$-axis [cf.\ equation~\eqref{eq:zetas}].

A second difference is that the proportionality constants appearing in the analogues of the decay law~\eqref{eq:frescrit} involve nonuniversal amplitudes. In order to obtain universal Casimir amplitudes, one must 
split off such nonuniversal factors. As expounded in section~\ref{sec:Delunivratios}, this can be achieved in a  natural fashion by defining universal Casimir amplitudes $\Delta^{\mathrm{BC}}_{\perp,\|}$ as universal amplitude ratios.%
\footnote{The issue of universality was recently considered in some detail also for the Casimir amplitudes of \emph{weakly anisotropic} scale-invariant systems \cite{Doh09,DC09}. As can be seen from these references (see, in particular, \cite[p.\ 15--17]{DC09}), the weak anisotropy can be absorbed by a proper choice of variables. Such a transformation to a system with a standard square-gradient term in the action's bulk density will in general change the boundary interaction constants of the corresponding continuum model and, in cases of fully finite systems, also their shapes.}
For given perpendicular or parallel surface plane orientation, these Casimir amplitudes depend on gross surface properties such as boundary conditions (BC). In cases of free BC, the BC that hold asymptotically in the large-length-scale limit are associated with the RG fixed points governing the surface universality classes of the respective semi-infinite systems. Owing to the different scaling behaviors of distances along the $\alpha$- and $\beta$-directions, the Hamiltonians of  appropriate minimal continuum models involve distinct boundary contributions $\mathcal{L}_{j=1,2}^{\perp,\|}$ in the cases of perpendicular and parallel surface plane orientations. We introduced these models and discussed the associated mesoscopic BC in section~\ref{sec:Mod}. 

Unfortunately, full analyses of these models for general values of the surface interaction constants are not even available for semi-infinite systems \cite{DGR03,DRG03,DR04,Die05}. For the sake of keeping the technical difficulties manageable, we restricted ourselves to the study of asymptotic large-length-scale BC of the $(\mathrm{O,O})$ type when considering free BC, namely the Dirichlet BC~\eqref{eq:bcaspar} and the BC $\bm{\phi}=\partial_n\bm{\phi}=\bm{0}$ [equation~\eqref{eq:bcasperp}] for parallel and perpendicular surface-plane orientations, respectively.  In addition, we investigated the cases of  periodic BC (PBC), both for parallel and perpendicular surface-plane orientations. The technical difficulties we encountered turned out to be somewhat easier to handle for parallel surface-plane orientations. For this orientation, we were able to compute the Casimir amplitudes $\Delta^{(\mathrm{O,O})}_{\|,\mathrm{G}}(m,d,n)$ and $\Delta^{\mathrm{PBC}}_{\|,\mathrm{G}}(m,d,n)$ of the corresponding Gaussian models. The results are given in equation~\eqref{eq:DeltaparGauss}. According to equation~\eqref{eq:CPLPrel}, these quantities are related in a simple manner to their CP analogues $\Delta^{(\mathrm{O,O})}_{\mathrm{CP,G}}(D,n)$ and $\Delta^{\mathrm{PBC}}_{\mathrm{CP,G}}(D,n)$ at $D=d-m/2$ dimensions.  As we showed in~\ref{app:largen}, the analogous relationship, equation~\eqref{eq:DeltaparrelSM}, holds between the large-$n$ Casimir amplitudes $\Delta_{\|,\infty}^{\mathrm{BC}}(d,m)$ and $\Delta_{\mathrm{CP},\infty}^{\mathrm{BC}}(d-m/2)$ for $(\mathrm{O,O})$ and periodic $\mathrm{BC}$. Our numerically determined  exact value for $\Delta_{\|,\infty}^{\mathrm{PBC}}(3,1)$ is given in equation~\eqref{eq:largenpar}. 

We also investigated the Casimir amplitudes $\Delta_{\|}^{(\mathrm{O,O})}(d,m,n)$ and $\Delta_{\|}^{\mathrm{PBC}}(d,m,n)$ by means of RG-improved perturbation theory in $d=4+m/2-\epsilon$ dimensions. As expected from the analogy with the CP case, $\Delta_{\|}^{(\mathrm{O,O})}(d,m,n)$ turned out to have a Taylor series expansion in  $\epsilon$. To order $\epsilon$, it is given in equation~\eqref{eq:DeltaparDD}. Owing to the presence of a zero mode at zero-loop order,  the conventional RG-improved perturbation theory for $\Delta_{\|}^{\mathrm{PBC}}(d,m,n)$ becomes ill-defined at the bulk LP because of IR singularities. To cope with this problem, we proceeded along the lines followed  in the CP case \cite{DGS06,GD08,DG09} and in analogous  problems of finite-temperature crossovers near quantum critical points \cite{Sac97}, using RG-improved perturbation theory to construct an effective $(d-1)$-dimensional action for the zero-mode component $\bm{\varphi}(\bm{y})$ of the order-parameter field $\bm{\phi}(\bm{y},z)$. The resulting modified RG-improved perturbation theory showed that the small-$\epsilon$ expansion of $\Delta_{\|}^{\mathrm{PBC}}(d,m,n)$ becomes a fractional one of a form similar to that of its CP analogue $\Delta^{\mathrm{PBC}}_{\mathrm{CP}}(d,n)$. It involves, besides integer powers of $\epsilon$, also half-integers powers $\epsilon^{k/2}$ with $k=3,5\ldots$ that are modulated by powers  of $\ln \epsilon$ when $k>3$.

Performing analogous calculations for perpendicular orientation turned out to be technically considerably more demanding. For the sake of simplicity, we therefore restricted ourselves to a one-loop approximation when considering the case of the $(\mathrm{O,O})$ BC~\eqref{eq:bcasperp}. This yielded the exact analytical expressions~\eqref{eq:DeltaGperp}, \eqref{eq:meq1}, and \eqref{eq:meqd} for the Gaussian Casimir amplitudes  $\Delta^{(\mathrm{O,O})}_{\mathrm{G},\perp}(d,m,n)$ in terms of double or single integrals.  In the case of $\mathrm{PBC}$, we were able to perform a two-loop calculation, but were faced again with the problem that a zero mode is present in Landau theory for finite $L$ at the bulk LP.  To deal with it, we pursued the above-mentioned strategy of constructing an effective $(d-1)$-dimensional action for the zero-mode component of the order parameter by integrating out the remaining nonzero-mode degrees of freedom by means of RG-improved perturbation theory. Unfortunately, the mass term of the free propagator of the resulting effective action  turned out to become negative near the upper critical dimension in the physically interesting uniaxial case. Upon inspection of the $\epsilon$ dependence of the corresponding formal perturbation series of $\Delta^{\mathrm{PBC}}_{\perp}(d,m,n)$ we could identify terms involving fractional powers $\sim \epsilon^{7/4}$ and analogous fractional powers of higher orders times powers of $\ln\epsilon$ [see equations~\eqref{eq:DelperpPBCepsexp}--\eqref{eq:a7ov4}]. Owing to the mentioned behavior of the free propagator's mass term,  we refrained from attempts of extrapolating this series to $d=3$. (Such extrapolations would require reasonable assumptions about the resummation of the series.)  
On the other hand, we found that a well defined large-$n$ limit of $\Delta_{\perp}^{\mathrm{PBC}}(3,1,n)/n$ exists; the exact limiting value $\Delta_{\perp,\infty}^{\mathrm{PBC}}(3,1)$, which we determined by numerical means, is given in equation~\eqref{eq:Delperpprbinfty31}. Interestingly, this quantity is positive, so that the corresponding Casimir force is repulsive.
 
To put our results and the problems encountered in the applied RG-improved perturbation theory in perspective, it is appropriate to recall a general difficulty one is faced with in studies of films near bulk criticality in the limit of large film thickness $L$: ultimately, one would like to have a theory that can handle the involved crossovers from $d$-dimensional (multi)critical behavior to $(d-1)$-dimensional (multi)critical or pseudo-(multi)critical behavior. However, perturbative RG approaches that are capable of dealing with critical behavior in bulk and semi-infinite systems (such as those based on the $\epsilon$ expansion) fall short of reaching this ambitious goal. Even the more modest aim of obtaining well-defined asymptotic expansions in $\epsilon$ at the bulk (multi)critical point may prove elusive for certain large-distance BC involving zero modes at the level of Landau theory. It comes as no surprise that these problems turned out to be more pronounced and severe than in previous analogous studies of Casimir interactions at bulk critical points \cite{KD91,KD92a,DGS06,GD08,DG09}. The reason is that the present study of finite-size properties involves a bulk \emph{multi}critical point, the LP. If the values of $d$ and $m$ are such that a LP continues to exist for finite $L$, the theory must be able to account in a reliable fashion for the  shifts of the LP and the phase boundaries that meet at it. For $d=3$ and finite $L$, the situation is even more complicated. One expects the thermal fluctuations to destabilize the phases with homogeneous and modulated order so that the critical lines get replaced by pseudo-critical ones and no real LP may be expected to exist for finite $L$ (see \cite{Sel92, Die02,Sel88} and their references). 

An appealing feature of the large-$n$ limit is that it is capable of dealing with  thermal fluctuations and dimensional crossover in a mathematically controlled fashion. On the other hand, one must keep in mind that in the  large-$n$ analyses presented here we fixed the thermodynamic fields to their values at the bulk LP point and hence did not use them to investigate the phase diagram for finite $L$. The results are meaningful for the particular (large-$n$ or spherical) models to which they apply. However, caution is necessary in applying them to the case of finite $n$. To draw meaningful conclusions, knowledge about the phase diagram from other sources such as Monte Carlo simulations or alternative approaches is necessary. 

Finally, let us briefly comment on the question of verification of the Casimir amplitudes investigated here. Previous experimental investigations of thermodynamic Casimir forces have focused for very good reasons on fluid systems  \cite{GC99,HHGDB08,ML00,FYP05,RBM07}. The advantage of using fluid systems is that the number of degrees of freedom of the medium (fluid) between macroscopic bodies (e.g., walls) can vary, a fact which enables direct and indirect confirmations of such fluctuation-induced forces. By contrast, the continuum models exhibiting anisotropic scale invariance on large distance scales considered here have natural realizations as lattice models (such as the ANNNI model in the uniaxial case \cite{Sel92,Die02,Sel88}) and rely on the presence of corresponding lattice anisotropies. Unlike fluid systems, such systems do not lend themselves to  direct  measurements of thermodynamic Casimir forces. On the other hand, the Casimir amplitudes we considered are finite-size quantities. 
Just as other finite-size quantities (which are not necessarily measurable via induced forces) , they are well-defined observables that at least in principle should be measurable. At the moment, the most promising 
means of checking our predictions appears to be Monte Carlo simulations. This technique was successfully used already some time ago to study the  surface critical behavior of bounded three-dimensional ANNNI models \cite{Ple02}. Appropriate extensions of this work along the lines of recent Monte Carlo investigations of thermodynamic Casimir forces near critical points  \cite{Huc07,VGMD07,Has10}  should be possible and enable reliable checks of our predictions. We hope that the present work will stimulate such investigations of fluctuation-induced interactions in strongly anisotropic critical systems.

\ack 
MASh thanks H.~W.\ Diehl and Fakult\"at f\"ur Physik for their hospitality at the
Universit\"at Duisburg-Essen. Partial support by DFG under grant no.~Di 378/5 is gratefully acknowledged.
\appendix

\section{Feynman graphs for the case of parallel slab orientation}\label{app:fg}

In this appendix we compute the one-loop and two-loop free-energy Feynman graphs that are needed to obtain our results for parallel slab orientation presented in section~\ref{sec:calcpar}.

To determine the one-loop graph in equation~\eqref{eq:fpsigraphs}, we set  $\tb=\rhob=0$, evaluate the required trace in the eigenfunction representation, and then exploit relation~\eqref{eq:Cmrel} to get rid of the quartic $k$-dependence. This gives
\begin{eqnarray}
\label{eq:oneloopfpsi}
\raisebox{-8pt}{\includegraphics[width=22pt,clip]{./oneloopfpsi.eps}}/A&=&
\frac{-1}{2A}\Tr'\ln\left(\frac{\delta^2\mathcal{H}[\bm{\psi}]}{\delta\psi_{a}(\bm{x})\delta\psi_{a'}(\bm{x}')}\bigg|_{\bm{\psi}=\bm{0}}\right)\nonumber\\
&=&\frac{-n}{2}\int_{\bm{k}}^{(m)}\int_{\bm{p}}^{(d-m-1)}\sum_{P_r\ne 0}\ln\left(p^2+P_r^2+\sigb k^4
\right)\nonumber\\
&=&\frac{-n}{2}\,\mathring{\sigma}^{-m/4}\,C_m\int_{\bm{q}}^{(d-m/2-1)}\sum_{P_r\ne 0}^\infty\ln\left[q^2+P_r^2
\right],
\end{eqnarray}
where $P_{r}$ are the BC-dependent momenta specified in equation \eqref{eq:varphiBC}. The prime on $\Tr$ reminds us that the trace in the case of PBC is restricted to the subspace orthogonal to the $P_r=0$ mode.

The integrals over the series in the last line are known from \cite{KD92a}, \cite{GD08}, and \cite{Sym81}. Using these results one arrives at
\begin{eqnarray}
\label{eq:oneloopfpsi2}
\raisebox{-8pt}{\includegraphics[width=22pt,clip]{./oneloopfpsi.eps}}/A&=&-\sigb^{-m/4}C_m\Big[nL\,X_{\mathrm{b}}(d-m/2)+nX^{\mathrm{BC}}_{\mathrm{s}}(d-m/2)\nonumber\\ &&\strut+\Delta^{\mathrm{BC}}_{\mathrm{CP},\mathrm{G}}(d-m/2)\,L^{-(d-1-m/2)}\Big],
\end{eqnarray}
where $\Delta^{\mathrm{BC}}_{\mathrm{CP},\mathrm{G}}(D)$ is the CP Casimir amplitude   of equation~\eqref{eq:CPLPrel} at dimension $D\equiv d-m/2$. Further,  $X_{\mathrm{b}}(D)$ ($=\int_{\bm{k}}^{(D)}\ln k^2$) and $X^{\mathrm{BC}}_{\mathrm{s}}(D)$ are $L$-independent functions of $D$ that vanish in dimensional regularization but would diverge  as $\Lambda^D$ and $ \Lambda^{D-1}$, respectively,  if a large-momentum cutoff $\Lambda$ were used to regularize the UV singularities. Both terms do not contribute to the residual free energy density. The $L$-dependent  last term yields the contribution to $f^{\mathrm{BC}}_{\mathrm{res}}(L;0,0,0,\sigma,\mu)$ given in equations~\eqref{eq:fresparGauss} and \eqref{eq:DeltaparGauss}.

The two-loop term in equation~\eqref{eq:fpsigraphs} reads
\begin{equation}
\label{eq:fpsi2graph}
\raisebox{-14.75pt}{\includegraphics[width=18pt,clip]{./twoloopfpsi.eps}}\,/A=\frac{n(n+2)}{3}\,\frac{\mathring{u}}{8}\,\int_0^L\rmd{z}\,{\left[G^{\mathrm{BC}}_{\psi,L;\|}(\bm{x};\bm{x})\right]}^2.
\end{equation}
Upon substituting the eigenfunction representation of the propagator $G^{\mathrm{BC}}_{\psi,L;\|}$, shown in the first line of equation~\eqref{eq:Gpsi}, we can once more harness relation~\eqref{eq:Cmrel} to conclude that this graph is given by its CP analogue in $D=d-m/2$ dimension times the factor $\sigma^{-m/2}C_m^2$. Insertion of the results for the corresponding CP graphs given in the literature \cite{KD92a,GD08} then yields
\begin{equation}
\label{eq:fpsi2graphres}
\raisebox{-14.75pt}{\includegraphics[width=18pt,clip]{./twoloopfpsi.eps}}\,/A=\frac{n(n+2)}{3}\,\frac{\mathring{u}}{8}\,\frac{C_m^2}{\sigb^{m/2}}L^{5-2d+m}\,I_2^{\mathrm{BC}}(d-m/2),
\end{equation}
where $I_2^{\mathrm{BC}}(D)$ stands for
\begin{equation}
\label{eq:IDPBC}
I_2^{\mathrm{PBC}}(D)=\left[ 2^{-1}\pi ^{-D/2} \zeta(D-2) \Gamma(D/2-1) \right]^2
\end{equation}
or
\begin{equation}
\label{eq:IDord}
I_2^{(\mathrm{O},\mathrm{O})}(D)=\frac{2^{2-2D}\pi^{D-3}[2\,\zeta^2(3-D)+\zeta(6-D)]}{\Gamma^2[(D-1)/2]\cos^2(D\pi/2)},
\end{equation}
depending on the BC.

Insertion of these results into equation~\eqref{eq:fpsigraphs} gives
\begin{eqnarray}\fl
\label{eq:fpsiBC}
f^{\mathrm{BC}}_{\psi,\mathrm{res}}(L)=\frac{C_m}{\sigb^{m/4}}\,\Delta^{\mathrm{BC}}_{\mathrm{CP},\mathrm{G}}(d-m/2,n)\,L^{-d+m/2+1}\nonumber \\
\,+\frac{n(n+2)}{3}\,\frac{\mathring{u}}{8}\,\frac{C_m^2}{\sigb^{m/2}}I_2^{\mathrm{BC}}(d-m/2)\,L^{-2d+m+5},\quad \mathrm{BC= PBC, (O,O)}\,,\nonumber\\
\end{eqnarray}
where $\Delta^{\mathrm{BC}}_{\mathrm{CP},\mathrm{G}}(D,n)$ are the Gaussian CP Casimir amplitudes
\begin{equation}\label{eq:DeltaPBCCP}
\Delta^{\mathrm{PBC}}_{\mathrm{CP},\mathrm{G}}(D,n)=2^D \Delta^{\mathrm{(O,O)}}_{\mathrm{CP},\mathrm{G}}(D,n)=-n\,\frac{\Gamma(D/2)\,\zeta(D)}{\pi^{D/2}}.
\end{equation}

\section{Large-\texorpdfstring{$n$}{n} limit}\label{app:largen}

The purpose of this appendix is to prove that the relation~\eqref{eq:CPLPrel} between the Gaussian LP Casimir amplitudes $\Delta_{\|,\mathrm{G}}^{\mathrm{BC}}$  and their CP analogues $\Delta_{\mathrm{CP},\mathrm{G}}^{\mathrm{BC}}$ carries over to the $\mathring{u}\ne 0$ theory in the limit $n\to\infty$, as stated in equation~\eqref{eq:DeltaparrelSM}.

To study the models introduced in section~\ref{sec:Mod} in the large-$n$ limit, we
rescale the coupling constant  as in \eqref{eq:gdef} and consider the partition function
\begin{equation}
\label{eq:Zndef}
\mathcal{Z}_n=\exp\left[-\frac{F_n}{k_BT}\right]\equiv\int\mathcal{D}[\bm{\phi}]\,\rme^{-\mathcal{H}[\bm{\phi}]}.
\end{equation}
Using the Einstein-like conventions that pairs of indices $a$, $\alpha$ and $\beta$ are to be summed over the values $a=1,\ldots,n$, $\alpha=1,\ldots,m$, and $\beta=m+1,\ldots,d$, respectively, we write the Hamiltonian as
\begin{equation}
\label{eq:Hrewr}
\mathcal{H}=\int_{\mathfrak{V}}\left[\frac{1}{2}\phi_a(\mathcal{K}-\rhob\,\partial_\alpha^2+\mathring{\tau}
)\phi_a+\frac{\mathring{g}}{4!n}\phi^4\right]
\end{equation}
where $\mathcal{K}$ means the differential operator
\begin{equation}
\mathcal{K}=\sigb \left(\partial^2_\alpha\right)^2-\partial_\beta^2
\end{equation}
subject to the boundary conditions discussed in section~\ref{sec:BC}. We do not have to specify these at this stage but will do so later. 

Following a standard technique \cite{Eme75}, we introduce the energy density operator $\mathcal{E}(\bm{x})=\phi^2(\bm{x})$ and rewrite the interacting part at a given position $\bm{x}$ as
\begin{eqnarray}
\label{eq:psiint}
\exp\left[-\frac{\mathring{g}}{n4!}\,\phi^4\right]&=&n\int_{-\infty}^\infty \rmd{\mathcal{E}}\,\delta(n\mathcal{E}-\phi_a\phi_a/2)\,\exp\left[-\frac{\mathring{g}n}{6}\,\mathcal{E}^2\right]\nonumber\\ &=&\frac{n}{2\pi\rmi}\int_{-\rmi\infty}^{\rmi\infty}\rmd{\vartheta}\int_{-\infty}^\infty \rmd{\mathcal{E}}\,\exp\bigg[\vartheta(n\mathcal{E}-\phi^2/2)-\frac{n\mathring{g}}{6}\,\mathcal{E}^2\bigg].\nonumber\\
\end{eqnarray}
We insert this expression into the partition function~\eqref{eq:Zndef} and then do the functional integral over $\bm{\phi}$ first. This is Gaussian, giving
\begin{eqnarray}
\lefteqn{\int\mathcal{D}[\bm{\phi}]\exp\bigg\{-\frac{1}{2}\int_{\mathfrak{V}}\left[\phi_a\big(\mathcal{K}+\tb+\vartheta-\rhob\,\partial_\alpha^2\big)\phi_a\right]\bigg\}}&&\nonumber\\&=&\exp\Big[ -\frac{n}{2}\ln\det\left(\mathcal{K}+\mathring{\tau}+\vartheta-\rhob\,\partial_\alpha^2\right )+\frac{n}{2}\,N_{\bm{x}}\ln(2\pi)\Big].
\end{eqnarray}
Here $N_{\bm{x}}=v^{-1}_0\int_{\mathfrak{V}}\rmd{V}$ is the number of lattice points $\bm{x}$, where $v_0$ denotes an elementary unit cell (discretization volume).
The partition function becomes
\begin{equation}
\label{eq:Znres}
\fl
\mathcal{Z}_n=C_n\int\mathcal{D}[\vartheta]\int\mathcal{D}[\mathcal{E}]\exp\Bigg\{n\Bigg[\int_{\mathfrak{V}} 
\left(\vartheta\mathcal{E}-\frac{g}{6}\,\mathcal{E}^2\right)-\frac{1}{2}\ln\det\big(\mathcal{K}+\mathring{\tau}+\vartheta-\rhob\,\partial_\alpha^2\big)\Bigg]\Bigg\}
\end{equation}
with $C_n=(n/\rmi)^{N_{\bm{x}}}(2\pi)^{(n-2)N_{\bm{x}}/2}$. To obtain the large-$n$ limit, we can evaluate the functional integrals by the method of steepest descent. The saddle-point equations become
\begin{eqnarray}
\label{eq:saddlept}
\vartheta(\bm{x})&=&\frac{\mathring{g}}{3}\,\mathcal{E}(\bm{x}),\\[\smallskipamount]
\label{eq:saddleptb}
\mathcal{E}(\bm{x})&=&\frac{1}{2}\,\big\langle\bm{x}\big|(\mathcal{K}+\mathring{\tau}+\vartheta-\rhob\,\partial_\alpha^2)^{-1}\big|\bm{x}\big\rangle.
\end{eqnarray}
Let $\mathcal{E}(\bm{x})$ and $\vartheta(\bm{x})$ be solutions to these equations that maximize the integrand in equation~\eqref{eq:Znres}. Then the large-$n$ limit of the reduced free energy per number $n$ of components is given by
\begin{equation}
\label{eq:Flargenres}
\lim_{n\to\infty}\frac{F_n}{n\,k_BT}=\frac{1}{2}\ln\det(\mathcal{K}+\mathring{\tau}+\vartheta-\rhob\,\partial_\alpha^2\big)-\int_{\mathfrak{V}}\Bigg[\frac{\mathring{g}}{6}\,\mathcal{E}^2+\frac{\ln( 2\pi)}{2v_0}\Bigg].
\end{equation}

For bulk systems and finite-size systems with PBC, the saddle-point equations~\eqref{eq:saddlept} have spatially homogeneous solutions $(\vartheta(\bm{x}),\mathcal{E}(\bm{x}))=
(\vartheta_{\mathrm{b}},\mathcal{E}_{\mathrm{b}})$ and  $(\vartheta^{\mathrm{PBC}}_{L},\mathcal{E}^{\mathrm{PBC}}_{L})$, respectively.  Furthermore, $r_{\mathrm{b}}=\vartheta_{\mathrm{b}}+\mathring{\tau}$ and
$r_{L}=\vartheta_{L}+\mathring{\tau}$ are the inverse bulk and finite-size susceptibilities. At the LP, both $r_{\mathrm{b}}$ and the $k^2=q_\alpha^2$ term of the Fourier transform of the bulk two-point vertex function $\mathcal{K}+r_{\mathrm{b}}-\rhob\,\partial_\alpha^2$ must vanish. Hence the critical values are given by
\begin{equation}
\rhob_{\mathrm{LP}}=0
\end{equation}
and
\begin{equation}
\label{eq:tauLP}
\mathring{\tau}_{\mathrm{LP}}=-\vartheta_{\mathrm{LP}}=-\frac{\mathring{g}}{6}\int^{(m)}_{\bm{k}}\int^{(d-m)}_{\bm{p}}\frac{1}{\sigb \,k^4+p^2}.
\end{equation}
If we continue to use dimensional regularization (assuming that $d>2+m/2$ to avoid IR singularities), we can use again equation~\eqref{eq:Cmrel} to relate $\mathring{\tau}_{\mathrm{LP}}$ to its CP analogue in $D=d-m/2$ dimensions:
\begin{eqnarray}
\label{eq:tauLPtauCP}
\mathring{\tau}_{\mathrm{LP}}(d,m)&=&\sigb^{-m/4}\,C_m\,\mathring{\tau}_{\mathrm{CP}}(d-m/2),\nonumber\\\mathring{\tau}_{\mathrm{CP}}(D)&=&-\frac{\mathring{g}}{6}\int_{\bm{q}}^{(D)}\frac{1}{q^2}.
\end{eqnarray}

We now set $(\mathring{\tau},\rhob)=(\mathring{\tau}_{\mathrm{LP}},\rhob_{\mathrm{LP}})$, introduce the deviations of $\vartheta$ and $\mathcal{E}$ from their bulk LP values,
\begin{equation}
 \label{eq:delthetae}
\delta\vartheta(\bm{x})=\vartheta(\bm{x})-\vartheta_{\mathrm{LP}},\quad \delta\mathcal{E}(\bm{x})=\mathcal{E}(\bm{x})-\mathcal{E}_{\mathrm{LP}},
\end{equation}
and consider the case of parallel orientation for PBC. Looking for spatially homogeneous solutions $\delta\vartheta^{\mathrm{PBC}}_{\|,L}(d,m;\mathring{g})$  and $\delta\mathcal{E}_{\|,L}^{\mathrm{PBC}}(d,m;\mathring{g})$, we find the self-consistent equations
\begin{eqnarray}
\label{eq:spteqPBC}\fl
\delta\vartheta^{\mathrm{PBC}}_{\|,L}(d,m;\mathring{g})=\frac{\mathring{g}}{3}\,\delta\mathcal{E}_{\|,L}^{\mathrm{PBC}}(d,m;\mathring{g})\nonumber\\ \fl =\frac{\mathring{g}}{6}\,\int_{\bm{k}}^{(m)}\int_{\bm{p}}^{(d-m-1)}\Bigg[\frac{1}{L}\sum_{r=-\infty}^\infty \frac{1}{p^2+(2\pi r/L)^2+\sigb\,k^4+\delta\vartheta^{\mathrm{PBC}}_{\|,L}(d,m;\mathring{g})}
\nonumber\\ \qquad\strut
 -\int_{-\infty}^\infty\frac{\rmd{P}}{2\pi}\,\frac{1}{p^2+P^2+\sigb\,k^4}\Bigg].
\end{eqnarray}

We can now apply relation~\eqref{eq:Cmrel} to the momentum integrals $\int^{(m)}_{\bm{k}}\int^{(d-m)}_{\bm{p}}$  of the function $h(p^2+\sigb\,k^4)$ inside the square brackets.  The resulting transformed equation agrees with its counterpart for the CP quantity $\delta\vartheta^{\mathrm{PBC}}_{\mathrm{CP},L}(D;\mathring{g}')$ of  a $(D=d-m/2)$-dimensional system with the coupling constant $\mathring{g}'=\mathring{g}\sigb^{-m/4}\,C_m$. Consequently, we have
\begin{equation}
\label{eq:thetaLPCPrel}
\delta\vartheta^{\mathrm{PBC}}_{\|,L}(d,m;\mathring{g})=\delta\vartheta^{\mathrm{PBC}}_{\mathrm{CP},L}(d-m/2;\mathring{g}\sigb^{-m/4}C_m)
\end{equation}
and
\begin{equation}
\label{eq:ELPCPrel}
\delta\mathcal{E}^{\mathrm{PBC}}_{\|,L}(d,m;\mathring{g})=\sigb^{-m/4}C_m\,\delta\mathcal{E}^{\mathrm{PBC}}_{\mathrm{CP},L}(d-m/2;\mathring{g}\sigb^{-m/4}C_m).
\end{equation}

In view of equations~\eqref{eq:tauLP} and \eqref{eq:tauLPtauCP}, it is clear that these relations carry over to $\vartheta^{\mathrm{PBC}}_{\|,L}(d,m;\mathring{g})$ and $\mathcal{E}^{\mathrm{PBC}}_{\|,L}(d,m;\mathring{g})$. Hence the term $\propto\mathcal{E}^2$ in equation~\eqref{eq:Flargenres} can be expressed in terms of its CP analogue as
\begin{equation}
\frac{\mathring{g}}{6}\,\left[\mathcal{E}_{\|,L}^{\mathrm{PBC}}(d,m;\mathring{g})\right]^2=\sigma^{-m/4}\,C_m\,\frac{\mathring{g}'}{6}\,\left[\mathcal{E}_{\mathrm{CP},L}^{\mathrm{PBC}}(D;\mathring{g}')\right]^2.
\end{equation}
That the Gaussian free energy term in equation~\eqref{eq:Flargenres} is related to its CP analogue in the same manner follows from our analysis of the Gaussian theory  in section~\ref{sec:calcpar} in conjunction with the results of \ref{app:fg} and relation~\eqref{eq:thetaLPCPrel}.  The upshot is that the $n=\infty$ Casimir amplitudes $\Delta_{\|,\infty}^{\mathrm{PBC}}$ are indeed related to their CP analogues  $\Delta_{\mathrm{CP},\infty}^{\mathrm{PBC}}$ as stated in equation~\eqref{eq:DeltaparrelSM}. This conclusion exploits the fact that the dependence  on the coupling constants $\mathring{g}$ and $\mathring{g}'$  drops out of these Casimir amplitudes because of their universality, albeit giving rise to  usual corrections to scaling.

Next, we turn to the case of parallel orientation and $(\mathrm{O,O})$ BC. Owing to the breakdown of translation invariance along the $z$-direction, the corresponding solutions $\vartheta^{\mathrm{(O,O)}}_{\|,L}$ and $\mathcal{E}^{\mathrm{(O,O)}}_{\|,L}$ to the saddle-point equations~\eqref{eq:saddlept}-\eqref{eq:saddleptb} become $z$-dependent. The equations for the shifted quantities $\delta\vartheta^{\mathrm{(O,O)}}_{\|,L}$ and $\delta\mathcal{E}^{\mathrm{(O,O)}}_{\|,L}$ now read
\begin{eqnarray}
\label{eq:spteqord}\fl
\lefteqn{\delta\vartheta^{\mathrm{(O,O)}}_{\|,L}(z|d,m;\mathring{g})=\frac{\mathring{g}}{3}\,\delta\mathcal{E}_{\|,L}^{\mathrm{(O,O)}}(z|d,m;\mathring{g})}&&\nonumber\\ &=&\frac{\mathring{g}}{6}\,\int_{\bm{k}}^{(m)}\int_{\bm{p}}^{(d-m-1)}\Bigg[\sum_{r} \frac{|\varphi_{r,L}(z;\mathring{g})|^2}{p^2+\varepsilon_{r,L}(\mathring{g})+\sigb\,k^4}\nonumber\\ &&\strut-\int_{-\infty}^\infty\frac{\rmd{P}}{2\pi}\,\frac{1}{p^2+P^2+\sigb\,k^4}\Bigg].
\end{eqnarray}
Here $\{\varphi_{r,L}\}$ is an orthonormal complete set of eigenfunctions on $[0,L]$ (which can be chosen real) and $\varepsilon_{r,L}$ are the corresponding eigenvalues. These functions satisfy
\begin{equation}
\left[-\partial_z^2+\vartheta_{\|,L}^{\mathrm{(O,O)}}(z;\mathring{g})-\varepsilon_{r,L}\right]\varphi_{r,L}(z;\mathring{g})=0
\end{equation}
and Dirichlet BC at $z=0$ and $z=L$. Their completeness and orthonormality relations read
\begin{equation}
\label{eq:completeness}
\sum_r\varphi^*_{r,L}(z;\mathring{g})\,\varphi_{r,L}(z';\mathring{g})=\delta(z-z')
\end{equation}
and 
\begin{equation}
\label{eq:orthonormality}
\int_0^L\rmd{z}\,\varphi^*_{r,L}(z;\mathring{g})\,\varphi_{r',L}(z;\mathring{g})=\delta_{rr'},
\end{equation}
respectively.

Obtaining analytical solutions to the above self-consistent equations is quite a challenge. Whether this goal can be achieved is unclear to us, both in the LP case we are concerned with here and the CP case with Dirichlet (or free) boundary conditions.  Bray and Moore \cite{BM77} managed to find analytical solutions for the CP quantity $\delta\vartheta_{\mathrm{CP},L=\infty}(z)$ in semi-infinite geometry. However, for finite $L$, one generally will have to resort to numerical techniques, as was done  in a recent large-$n$ study of the CP Casimir effect \cite{Comtesse09}. 

Numerical calculations of the eigensystem $\{\varepsilon_{r,L},\varphi_{r,L}\}$ require in one way or another a cutoff $\Lambda_z=\pi/a_z$, where   $a_z$ is a discretization length (lattice constant) for $z$. On the other hand, it is preferable to do the required momentum integrations $\int_{\bm{p}}^{(d-m-1)}\int_{\bm{k}}^{(m)}$ analytically as before, and not restrict them by a cutoff. The price is that UV singularities  are encountered in the free energy, even for $d$ strictly below the upper critical dimension $d^*(m)=4+m/2$ (where the theory is superrenormalizable). To obtain convergent integrals, one must (and can) make appropriate subtractions. (Subtracting the bulk and surface free energy contributions to obtain the excess free energy density will remove the bulk and surface UV singularities associated with additive counterterms. For $d<d^*(m)$, the resulting differences must be UV finite.) In the equations for the shifted variables, UV finiteness is ensured when $d<d^*(m)$ because of the subtracted bulk terms. 
Reasoning as in the case of $\mathrm{PBC}$, we can therefore conclude that the following analogues of relations~\eqref{eq:thetaLPCPrel} and \eqref{eq:ELPCPrel} hold:
\begin{equation}
\label{eq:thetaLPCPrelz}
\delta\vartheta^{\mathrm{(O,O)}}_{\|,L}(z|d,m;\mathring{g})=\delta\vartheta^{\mathrm{(O,O)}}_{\mathrm{CP},L}(z|d-m/2;\mathring{g}\sigb^{-m/4}C_m)
\end{equation}
and
\begin{equation}
\label{eq:ELPCPrelz}
\delta\mathcal{E}^{\mathrm{(O,O)}}_{\|,L}(z|d,m;\mathring{g})=\frac{C_m}{\sigb^{m/4}}\,\delta\mathcal{E}^{\mathrm{(O,O)}}_{\mathrm{CP},L}(z|d-m/2;\mathring{g}\,\sigb^{-m/4}C_m).
\end{equation}
The residual free-energy densities $f^{(\mathrm{(O,O)}}_{\|,\mathrm{LP}}(L|d,m;\mathring{g})$ and $f^{(\mathrm{(O,O)}}_{\mathrm{CP}}(L|D;\mathring{g}')$ must therefore be related in the same manner, which in turn implies relation~\eqref{eq:DeltaparrelSM} for the Casimir amplitudes.

\section{Feynman graphs for the case of perpendicular slab orientation and periodic boundary conditions} \label{app:fgperppbc}
\label{app:fgperp}

In this appendix we compute the Feynman graphs required for our calculation of the Casimir amplitudes for perpendicular orientation and PBC reported in \sref{sec:CAperpPRB}.

We begin with the calculation of the one-loop contribution  to the free-energy area density
\begin{equation}
  \label{eq:fperppbc1}
 f^{\mathrm{PBC}}_{\perp,[1]}(L|\sigb;\tb,\rhob)=\frac{n}{2}\int^{(d-m)}_{\bm{p}}\int_{\bm{k}}^{(m-1)}\sum_{r\in\mathbb{Z}}\ln\Omega_{\sigb}(\bm{q}_r|\tb,\rhob)
\end{equation}
for $\tb=\rhob=0$. Here 
\begin{equation}\label{eq:Omegaq}
\Omega_{\mathring{\sigma}}(\bm{q}_r|\mathring{\tau},\mathring{\rho})=\mathring{\tau}+p^2+\mathring{\rho}\,k^2+\sigb{\big[k^2+(2\pi r/L)^2\big]}^2
\end{equation}
and $\bm{q}_r$ denotes the $d$-dimensional momentum vector $(k_1,\ldots,k_{m-1},\frac{2\pi r}{L},p_1,\ldots,p_{d-m})$.

In the limit $L\to\infty$, the ratio $f^{\mathrm{PBC}}_{\perp,[1]}(L|\sigb;\tb,\rhob)/L$ approaches the  one-loop term  of the reduced bulk  free-energy density $f_{\mathrm{b}}(\sigb;\tb,\rhob)$, namely
\begin{equation}
 \label{eq:fb1}
 f_{\mathrm{b},[1]}(\sigb;\tb,\rhob) =\frac{n}{2}\int^{(d)}_{\bm{q}}\ln\Omega_{\sigb}(\bm{q}|\mathring{\tau},\rhob).
 \end{equation}

Subtracting from equation~\eqref{eq:fperppbc1} the bulk term and using Poisson's summation formula~\eqref{eq:Poisson} gives
\begin{eqnarray}
\label{eq:Delfperppbc1}
f^{\mathrm{PBC}}_{\mathrm{res},\perp;[1]}(L|\sigb;\tb,\rhob)&=& f^{\mathrm{PBC}}_{\perp;[1]}(L|\sigb;\tb,\rhob)-L f_{\mathrm{b},[1]}(\sigb;\tb,\rhob)\nonumber\\ &=&n\sum_{j=1}^\infty\int^{(d)}_{\bm{q}}\cos(q_m j L)\ln\Omega_{\sigb}(\bm{q}|\mathring{\tau},\rhob).
 \end{eqnarray}
 
 The summands in the second line of this equation may  be conveniently expressed in terms of the free bulk propagator $G_{\mathrm{b}}^{(d+2,m)}$. To show this, we insert the identity $1=\nabla_{\bm{p}}\cdot\bm{p}/(d-m)$ into the integral $\int_{\bm{p}}^{(d-m)}$ and integrate by parts, obtaining
  \begin{eqnarray}
  \label{eq:partint}
\int_{\bm{p}}^{(d-m)}\ln\Omega_{\sigb}(\bm{q}|\tb,\rhob)&=&- \int_{\bm{p}}^{(d-m)}\frac{\bm{p}\cdot\nabla_{\bm{p}}}{d-m}\,\ln\Omega_{\sigb}(\bm{q}|\tb,\rhob)\nonumber\\ &=&-4\pi \int_{\bm{P}}^{(d+2-m)}\frac{\partial}{\partial P^2}\ln\Omega_{\sigb}(\bm{k},\bm{P}|\tb,\rhob),
\end{eqnarray}
where $\bm{P}$ is a ($d+2-m$)-dimensional wave vector.
 We have exploited the facts that $\Omega_{\sigb}(\bm{q}|\tb,\rhob)$ depends  on $\bm{p}$  merely via $p^2$ and that $\bm{p}$-integrals over rotationally invariant functions $h(p^2)$ 
 simplify to
 \begin{equation}
 \int^{(D)}_{\bm{p}}h(p^2)=K_D\int_0^\infty \rmd{p}\,p^{D-1}h(p^2),
 \end{equation}
 with 
 \begin{equation}
K_D=2^{1-D}\pi^{-D/2}/\Gamma(D/2),
\end{equation}
the standard geometrical factor resulting from the angular integrations. 
 
The result~\eqref{eq:partint} suggests to introduce the functions
\begin{eqnarray}
\label{eq:Pdmdef}
P_{m,d}(\tau,\rho)&\equiv&\sum_{j=1}^\infty\int^{(d)}_{\bm{q}}\frac{2\cos(q_mj)}{\tau+\rho k^2+p^2+ k^4}\nonumber\\ &=&2\sum_{j=1}^\infty G^{(d,m)}_{\mathrm{b}}(j\hat{z}|1;\tau,\rho),
\end{eqnarray}
where $G^{(d,m)}_{\mathrm{b}}(\bm{x}|\sigb;\tb,\rhob)$ is the free bulk propagator~\eqref{eq:Gb}. It is straightforward to show that these functions have the property
\begin{equation}
\label{eq:Pprimerel}
\frac{\partial}{\partial\tau} P_{m,d+2}(\tau,\rho)= -\frac{1}{4\pi}\,P_{m,d}(\tau,\rho),
\end{equation}
which they share with the functions $Q_{d,2}(y)/y$ used in \cite{GD08} and \cite{DDG06} [cf.\ equations~(4.17) and (B5) of the first and second reference, respectively]. Let us differentiate the series of $P_{m,d+2}$ termwise with respect to $\tau$ and interchange the differentiation with the momentum integration $\int^{(d)}_{\bm{q}}$. The $\tau$-derivative of the integrand can be replaced by its derivative with respect to $p^2$. An integration by parts as in equation~\eqref{eq:partint} then gives relation~\eqref{eq:Pprimerel}.

Upon combining the above results, we finally arrive at
\begin{equation}
\label{eq:fresperpPRB}\fl
 f^{\mathrm{PBC}}_{\mathrm{res},\perp;[1]}(L|\sigb;\tb,\rhob)=-2\pi \,n\,\sigb^{(d-m)/2}\,L^{1-2d+m}\,P_{m,d+2}(\tb L^4/\sigb ,\rhob L^2/\sigb).
\end{equation}

In order to determine the Casimir amplitudes $\Delta_{\perp}^{\mathrm{PBC}}(d,m,n)$, we need the value of $P_{m,d+2}$ at $\tb=\rhob=0$, which we denote as $P_{m,d+2}^{(0)}$. To compute $P_{m,d}^{(0)}$, we  substitute the scaling form~\eqref{eq:GbLP} of the free bulk propagator into equation~\eqref{eq:Pdmdef} and use the asymptotic behavior~\eqref{eq:Phiasinfty} of the scaling function $\Phi_{m,d}$, obtaining
\begin{eqnarray}
   \label{eq:fperppsiPBC2}
  P_{m,d}^{(0)}\equiv  P_{m,d}(0,0)&=&2\sum_{j=1}^\infty G_{\mathrm{b}}^{(d,m)}(j\hat{\bm{z}}|1;0,0) \nonumber\\ &=&2\Phi_{m,d}^{(\infty)}\,\zeta(2d-m-4),
\end{eqnarray}
where $\Phi_{m,d}^{(\infty)}$ is the coefficient defined in equation~\eqref{eq:Phiinfty}. Its insertion into equation~\eqref{eq:fperppsiPBC2} yields the expression quoted in equation~\eqref{eq:Pmd0def}. From equation~\eqref{eq:fresperpPRB} one can read off  the expression for the  Gaussian Casimir amplitude $\Delta_{\perp,\mathrm{G}}^{\mathrm{PBC}}$ given in the first line of equation~\eqref{eq:DelperpPRBGauss}. The result presented in its second line follows upon insertion of the result for $P_{m,d+2}^{(0)}$ implied by equation~\eqref{eq:Pmd0def}. 


\end{document}